\documentclass[draftclsnofoot, onecolumn, article]{IEEEtran}
\usepackage{ifpdf}
\usepackage[noadjust]{cite}
\ifCLASSINFOpdf
   \usepackage[pdftex]{graphicx}
\else
\fi
\usepackage[cmex10]{amsmath}
\usepackage{amsfonts}

\usepackage{algorithmic}

\usepackage{array}
\usepackage{mdwmath}
\usepackage{mdwtab}
\usepackage{eqparbox}
\usepackage[font=footnotesize]{subfig}
\usepackage{amsfonts}
\newtheorem{theorem}{Theorem}

\newtheorem{lemma}[theorem]{Lemma}

\newtheorem{observation}[theorem]{Observation}

\newcommand{\field}[1]{\mathbb{#1}}
\newcommand{\R}{\field{R}} % real numbers
\renewcommand{\Re}{\R} % reals
\newcommand{\tends}{{\rightarrow}} % arrow for limits
\newcommand{\E}{\field{E}} % expectation 2
\newcommand{\Nscr}{{\cal N}}

\newcommand{\Pscr}{{\cal P}}

\newcommand{\Xscr}{{\cal X}}

\newcommand{\tsnr}{{\text{snr}}}
\newcommand{\tvar}{{\text{Var}}}
\newcommand{\tmmse}{{\text{mmse}}}
\newcommand{\tcmmse}{{\text{cmmse}}}

\newcommand{\lmmse}{{\text{lmmse}}}
\newcommand{\Ytil}{{\tilde{Y}}}

\newcommand{\bea}{\begin{eqnarray}}
\newcommand{\eea}{\end{eqnarray}}
\newcommand{\beas}{\begin{eqnarray*}}
\newcommand{\eeas}{\end{eqnarray*}}
\newcommand{\twopartdef}[4]
{
	\left\{
		\begin{array}{ll}
			#1 & \mbox{if } #2 \\
			#3 & \mbox{if } #4
		\end{array}
	\right.
}

\begin{document}
\title{Information, Estimation, and Lookahead in the Gaussian channel}
\author{Kartik Venkat\IEEEauthorrefmark{1}, Tsachy Weissman\IEEEauthorrefmark{1}, Yair Carmon\IEEEauthorrefmark{2}, Shlomo Shamai\IEEEauthorrefmark{2}
\thanks{\IEEEauthorrefmark{1} Stanford University \IEEEauthorrefmark{2} Technion, Israel Institute of Technology} \thanks{Part of this work was presented at the 50$^{th}$ Annual Allerton Conference on Communications, Control and Computing, Monticello, Illinois, 2012.}}
\maketitle

\begin{abstract}
%\boldmath

We consider mean squared estimation with lookahead of a continuous-time signal corrupted by additive white Gaussian noise. We show that the mutual information rate function, i.e., the mutual information rate as function of the signal-to-noise ratio (SNR), does not, in general, determine the minimum mean squared error (MMSE) with fixed finite lookahead, in contrast to the special cases with 0 and infinite lookahead (filtering and smoothing errors), respectively, which were previously established in the literature. We also establish a new expectation identity under a
generalized observation model where the Gaussian channel has an SNR jump at $t=0$, capturing the tradeoff between lookahead and SNR.

Further, we study the class of continuous-time stationary Gauss-Markov processes (Ornstein-Uhlenbeck processes) as channel inputs, and
explicitly characterize the behavior of the minimum mean squared error (MMSE) with finite lookahead  and signal-to-noise ratio (SNR). The MMSE with lookahead is shown to converge exponentially rapidly to the non-causal error, with the exponent being the reciprocal of the non-causal error. We extend our results to mixtures of Ornstein-Uhlenbeck processes, and use the insight gained to present lower and upper bounds on the MMSE with lookahead for a class of stationary Gaussian input processes, whose spectrum can be expressed as a mixture of Ornstein-Uhlenbeck spectra. 


\end{abstract}


% \fi
\IEEEpeerreviewmaketitle

\section{Introduction} \label{sec: introduction}
Mean squared estimation of a signal in the presence of Gaussian noise has been a topic of considerable importance to the communities of information theory and estimation theory. There have further been discoveries tying the two fields together, through identities between fundamental informational quantities  and the squared estimation loss.

Consider the continuous-time Gaussian channel. In \cite{duncan}, Duncan established the equivalence between the input-output mutual information and the integral of half the causal mean squared error in estimating the signal based on the observed process. In \cite{gsv2005}, Guo et al. present what is now known as the I-MMSE relationship, which equates the derivative of the mutual information rate to half the average non-causal squared error. These results were extended to incorporate mismatch at the decoder for the continuous-time Gaussian channel by Weissman in \cite{weissman10}, building upon similar relations for the scalar Gaussian channel by Verd\'u in \cite{ver_mis10}. In \cite{weissman10}, it was shown that when the decoder assumes an incorrect law for the input process, the difference in mismatched filtering loss is simply given by the relative entropy between the true and incorrect output distributions. Recently, in \cite{VenkatWeissman11}, a unified view of the aforementioned identities was presented. In particular, pointwise analogues of these and related results were developed, characterizing the original relationships as identities involving expectations of random objects with information-estimation significance of their own. 

In this work, we shall focus on stationary continuous-time signals corrupted by Gaussian noise. Let $\mathbf{X} = \{X_t, t \in \Re\}$ denote a
continuous-time signal, which is the channel input process. We assume square integrability of the stochastic process $\mathbf{X}$, which says that for any finite interval $[a,b] \subset \Re$, $\E [ \int_a^b X^2_t \, dt] < \infty$. 

The output process $\mathbf{Y}$ of the continuous-time Gaussian channel with input $\mathbf{X}$ is given by the following model 
\begin{equation}
dY_t = \sqrt{\tsnr}\,X_t dt + dW_t, \label{eq: channel model}
\end{equation}
for all $t$, where $\tsnr > 0$ is the channel signal-to-noise ratio (SNR) and $W_\cdot$ is a standard Brownian Motion (cf. \cite{KaratzasShreve} for definition and properties of Brownian motion) independent of $\mathbf{X}$.
For simplicity let us assume the input process to be stationary. Throughout, we will denote $X_a^b = \{X_t : a \leq t \leq b\}$. 

Let $I(\tsnr)$ denote the mutual information rate, given by
\bea
I(\tsnr) = \lim_{T \to \infty} \frac{1}{T} \,I(X_0^T; Y_0^T).
\eea
Let $\tmmse(\tsnr)$ denote the smoothing squared error 
\bea
\tmmse(\tsnr) = \lim_{T \to \infty} \frac{1}{T} \int_0^T \E[(X_t - \E[X_t |Y_{0}^{T}])^2] \, dt, \label{eq: average mmse}
\eea
which, in our setting of stationarity, can equivalently be defined as
\bea
\tmmse(\tsnr) = \E[(X_0 - \E [X_0 | Y_{-\infty}^{+\infty}])^2]. \label{eq: mmse for stationary process}
\eea
Similarly, the causal squared error is given by
\bea
\tcmmse(\tsnr) &=& \lim_{T \to \infty} \frac{1}{T} \int_0^T \E[(X_t - \E[X_t |Y_{0}^{t}])^2]\, dt \\
&=& \E[(X_0 - \E[X_0 | Y_{-\infty}^0])^2]. \label{eq: cmmse for stationary process}
\eea
From \cite{duncan} and \cite{gsv2005}, we know that the above quantities are related according to
\bea
\frac{2I(\tsnr)}{\tsnr} = \tcmmse(\tsnr) = \frac{1}{\tsnr}\int_0^{\tsnr}
\tmmse(\gamma) \,d\gamma \label{eq: I-cmmse-mmse relationship}
\eea
for all signal-to-noise ratios, $\tsnr > 0$, and for all choices of the underlying input distribution\footnote{Indeed, (\ref{eq: I-cmmse-mmse relationship}) continues to hold in the absence of stationarity of the input process, in which case the respective quantities are averaged over a fixed, finite interval $[0,T]$.} for which the process $\mathbf{X}$ is square integrable. Thus, the mutual information, the causal error and the non-causal error are linked together in a distribution independent fashion for the Gaussian channel under mean-squared loss. Having understood how the time averaged causal and non-causal errors behave, and in particular the relation (\ref{eq: I-cmmse-mmse relationship}) implying that they are completely characterized by the mutual information rate function - the current work seeks to address the question - what happens for lookahead between $0$ and $\infty$?.

The mean squared error with finite lookahead $d \geq 0$ is defined as
\bea
\lmmse(d,\tsnr) = \E[(X_0 - \E[X_0 | Y_{-\infty}^d])^2]  \label{eq: lmmse for stationary process},
\eea
where it is instructive to note that the special cases $d=0$ and $d=\infty$ in
(\ref{eq: lmmse for stationary process}) yield the filtering (\ref{eq: cmmse for stationary process}) and smoothing (\ref{eq: mmse for stationary process})
errors respectively. Let us be slightly adventurous, and rewrite (\ref{eq: I-cmmse-mmse relationship}) as
\bea
\frac{2I(\tsnr)}{\tsnr} = \E[\lmmse(0,\Gamma_0)] = \E[\lmmse(\infty,\Gamma_\infty)],
\eea
where the random variables $\Gamma_0, \Gamma_\infty$ are distributed according to $\Gamma_ 0 = \tsnr$ a.s., and $\Gamma_\infty \sim U[0,\tsnr]$, where $U[a,b]$ denotes the uniform distribution on $[a,b]$. This characterization of the mutual information rate may lead one to conjecture the existence of a random variable $\Gamma_d$, whose distribution depends only on $d, \tsnr$, and some features of the process (such as its bandwidth and spectrum), and satisfies the identity
\bea
\frac{2I(\tsnr)}{\tsnr} = \E[\lmmse(d,\Gamma_d)], \label{eq: conjecture}
\eea
regardless of the distribution of $\mathbf{X}$. I.e., there exists a family of distributions $\Gamma_d$ ranging from one extreme of a point mass at $\tsnr$ for $d=0$, to the uniform distribution over $[0,\tsnr]$ when $d=\infty$. One of the corollaries of our results in this work is that such a relation cannot hold in general, even when allowing the
distribution of $\Gamma_d$ to depend on distribution-dependent
features such as the process spectrum. In fact, in Section \ref{sec: mutual information and lmmse} we show that in general, the mutual information rate function of a process is not sufficient to characterize the estimation error with finite lookahead. Therefore, one would need to consider features of the process that go beyond the mutual information rate and the spectrum - in order to characterize the finite lookahead estimation error. Motivated by this question, however, in Section \ref{sec: generalized
observation model} we establish an identity which relates the filtering error to a double integral of the mean squared error, over lookahead and SNR, for the Gaussian channel with an SNR jump at $t=0$.

In the literature, finite lookahead estimation has been referred to as fixed-lag smoothing (cf., for example, \cite{Lipster_fixedlag} and references therein).  Note that (\ref{eq: lmmse for stationary process}) is well defined for all $d \in \R$. For $d < 0$, $\lmmse(d,\tsnr)$ denotes the finite horizon average prediction mean square error, which is as meaningful as its finite lookahead counterpart, i.e. the case when $ d > 0$. Motivated by the desire to understand the tradeoff between lookahead and mmse, in Section \ref{sec: Ornstein-Uhlenbeck Process} we explicitly characterize this trade-off for the canonical family of continuous-time stationary Gauss-Markov (Ornstein-Uhlenbeck) processes. In particular, we find that the rate of convergence of $\lmmse(\cdot,\tsnr)$ with increasing lookahead (from causal to the non-causal error) is exponential, with the exponent given by the inverse of the non-causal error, i.e. $\frac{1}{\tmmse(\tsnr)}$ itself.

The rest of the paper is organized as follows. In Section \ref{sec: Ornstein-Uhlenbeck Process} we explicitly characterize the lookahead-MSE trade-off for the canonical family of continuous-time stationary Gauss-Markov (Ornstein-Uhlenbeck) processes. In particular, we find that the convergence of $\lmmse(\cdot,\tsnr)$ with increasing lookahead (from causal to the non-causal error) is exponentially fast, with the exponent given by the inverse of the non-causal error, i.e. $\frac{1}{\tmmse(\tsnr)}$ itself. In Section \ref{sec: mixture of Ornstein-Uhlenbeck processes}, we extend our results to a larger class of processes, that are expressible as a mixture of Ornstein-Uhlenbeck processes. We then consider stationary Gaussian input processes in Section \ref{sec: stationary gaussian processes}, and characterize the MMSE with lookahead via spectral factorization methods. In Section \ref{sec: information utility}, we consider the information utility of infinitesimal lookahead for a general stationary input process and relate it to the squared filtering error. In Section \ref{sec: generalized
observation model}, we introduce a generalized observation model for a
stationary continuous-time signal where the channel has a non-stationary SNR
jump at $t=0$. For this model, we establish an identity relating the
squared error with finite lookahead and the causal filtering error. In Section \ref{sec: mutual information and lmmse}, we show by means of an example that a distribution-independent identity of the form in (\ref{eq: conjecture}) cannot hold in general. We conclude with a summary of our findings in Section \ref{sec: conclusions}.

\section{Ornstein-Uhlenbeck Process} \label{sec: Ornstein-Uhlenbeck Process}
\subsection{Definitions and Properties}
The Ornstein-Uhlenbeck process \cite{OrnsteinUlhenbeck30} is a classical continuous-time
stochastic process characterized by the following stochastic differential
equation
\begin{eqnarray}
dX_t = \alpha(\mu - X_t)dt + {\beta}dB_t, \label{eq: Ornstein-Uhlenbeck process evolution}
\end{eqnarray}
where $\{B_t\}_{t \geq 0}$ is a Standard Brownian Motion and $\alpha$, $\mu$, $\beta$ are process parameters. The reader is referred to \cite{Doob42} for a historical perspective on this process. Below, we present some of its important statistical properties. 
The mean and covariance functions are given by:
\begin{equation}
\E(X_t) = \mu, 
\end{equation}
and,
\begin{align}
\text{Cov}(X_t,X_s) &= \frac{\beta^{2}}{2\alpha}e^{-\alpha|t-s|}.
\end{align}
We can further denote the autocorrelation function and power spectral density of this process by
\begin{align}
R_{X}(\tau) &= \frac{\beta^{2}}{2\alpha}e^{-\alpha|\tau|}, \label{eq: autocorrelation of Ornstein-Uhlenbeck process}\\
S_{X}(\omega) &= \frac{\beta^{2}}{\alpha^2 + \omega^2}.  \label{eq: PSD of Ornstein-Uhlenbeck process}
\end{align}
%For fixed 0 $\leq$ s $\leq$ t, the random variable $X_t$ conditioned upon $X_s$ has a Gaussian distribution with mean and variance given by
%\begin{equation}
%E(X_t|X_s) = \mu + (X_s - \mu)e^{-\alpha(t-s)}, \hspace*{2em} \tvar(X_t|X_s) = \frac{\beta^{2}}{2\alpha}(1-e^{-2\alpha(t-s)}).
%\end{equation}
In all further analysis, we consider the process mean $\mu$ to be 0. We also note that all expressions are provided assuming $\alpha > 0$ which results in a mean-reverting evolution of the process.
\subsection{Mean Squared Estimation Error}

Recall the mean squared error with finite lookahead $d$ defined in (\ref{eq: lmmse for stationary process}) from which one can infer the filtering (\ref{eq: cmmse for stationary process}) and smoothing (\ref{eq: mmse for stationary process})
errors respectively. We now compute these quantities for the Ornstein-Uhlenbeck
process at a fixed signal to noise ratio $\gamma$. Define, for the
Ornstein-Uhlenbeck process corrupted by the Gaussian channel (\ref{eq: channel model}), the error in estimating $X_0$ based on a finite lookahead $d$ into the future and a certain lookback $l$ into the past:
\begin{eqnarray} \label{eq: nu for Ornstein-Uhlenbeck process}
\nu(l,d,\gamma) = \tvar(X_0|Y_{-l}^{d}) \hspace*{1em} d,l \geq 0.
\end{eqnarray}
Before we explicitly characterize $\nu(l,d,\gamma)$, we present the following
intermediate result, which is proved in Appendix \ref{appendix: Lemma 1}.
\begin{lemma} \label{lemma: ed for Ornstein-Uhlenbeck process}
Let $e_d = \nu(0,d,\gamma) = \nu(d,0,\gamma)$\footnote{where the latter equality holds due to time-reversibility of the Ornstein-Uhlenbeck process}. Then,
\begin{equation}
e_d = \frac{\Lambda_d \sqrt{\alpha^2 + \gamma{\beta^2}} - \alpha}{\gamma},
\end{equation}
where $\Lambda_d = \frac{e^{2d\sqrt{\alpha^2 + \gamma{\beta^2}}}\rho
+ 1}{e^{2d\sqrt{\alpha^2 + \gamma{\beta^2}}}\rho - 1}$ and $\rho =
\left|\frac{\gamma R_X(0) + {\sqrt{\alpha^2 + \gamma{\beta^2}} + \alpha}}{\gamma
R_X(0) - \sqrt{\alpha^2 + \gamma{\beta^2}} + \alpha}\right|$.
\end{lemma}
Using Lemma \ref{lemma: ed for Ornstein-Uhlenbeck process}, we now compute the more general
quantity $\nu(l,d,\gamma)$ defined in (\ref{eq: nu for Ornstein-Uhlenbeck process}), which
denotes the loss in estimating $X_t$ based on the observations
$Y_{t-d}^{t+l}$, for a stationary process.
\begin{lemma}[estimation error with finite window of observations in the past
and future]
\label{lemma: nu(l,d,gamma) for Ornstein-Uhlenbeck}
\bea
\nu(l,d,\gamma) = \frac{e_l e_d R_X(0)}{R_X(0)(e_l + e_d) - e_l e_d}. \label{eq: look ahead and back}
\eea
\end{lemma}
The proof of Lemma \ref{lemma: nu(l,d,gamma) for Ornstein-Uhlenbeck} is straightforward, upon noting (i) the Markov Chain relationship $Y_{-l}^0\--X_0\--{Y_0^d}$, and (ii) the joint Gaussianity  of ($X_0,Y_{-l}^d$). Thus, one can use the conditional variances $\tvar(X_0 | Y_{-l}^0)$ and $\tvar(X_0|Y_0^d)$ to obtain the required expression for $\tvar(X_0 | Y_{-l}^d)$.   
%\begin{IEEEproof}
%We note the following Markov chain relationship for all $l,d > 0$,
%\bea
%Y_{-l}^0\--X_0\--{Y_0^d}. \label{eq: markov relationship for Ornstein-Uhlenbeck}
%\eea
%We further note that ($X_0,Y_{-l}^d$) is jointly Gaussian. In particular, we
%know that
%\begin{align}
%\tvar(X_0) &= R_X(0), \label{eq: variances with lookahead and lookback 1}\\
%\tvar(X_0|Y_{-l}^0) &= e_l,\label{eq: variances with lookahead and lookback
%2}\\ \tvar(X_0|Y_0^d) &= e_d. \label{eq: variances with lookahead and lookback
%3}
%\end{align}
%Let $f_{X_t}(\cdot)$ denote the probability density function of $X_t$. Using the Markov relationship in (\ref{eq: markov relationship for Ornstein-Uhlenbeck}), it can be shown that,
%\bea
%f_{X_0|Y_{-l}^d}(x|y_{-l}^{d}) &=&
%\frac{f_{X_0|Y_{-l}^0}(x|y_{-l}^0)f_{X_0|Y_{0}^d}(x|y_{0}^d)}{\int_{-\infty}^{
%\infty}f_{X_0|Y_{-l}^0}(\tilde{x}|y_{-l}^0)f_{X_0|Y_{0}^d}(\tilde{x}|y_{0}^d)\,
%d\tilde{x}}.  \label{eq: pdf given (l,d)}
%\eea
%Further, since the random variables are jointly Gaussian, even the conditional
%distributions in  (\ref{eq: pdf given (l,d)}) are Gaussian, with variances as
%defined in (\ref{eq: variances with lookahead and lookback 1})-({\ref{eq:
%variances with lookahead and lookback 3}}), we can easily verify that the variance
%of $X_0|Y_{-l}^d$ is given by (\ref{eq: look ahead and back}).
%\end{IEEEproof}

Using Lemma \ref{lemma: ed for Ornstein-Uhlenbeck process} and Lemma \ref{lemma: nu(l,d,gamma)
for Ornstein-Uhlenbeck} we explicitly characterize the filtering (\ref{eq: cmmse for
stationary process}) and smoothing (\ref{eq: mmse for stationary process})
(mean squared error) MSE's for the noise corrupted Ornstein-Uhlenbeck process, in the following Lemma.
\begin{lemma}[MSE for the Ornstein-Ulhenbeck process] \label{lemma: mse OU}
For the Ornstein-Ulhenbeck process corrupted by white gaussian noise at signal-to-noise ratio $\tsnr$, the filtering and smoothing errors are given by:
\begin{itemize}
\item Filtering Error
\bea
\tcmmse(\tsnr) = \nu(\infty,0,\tsnr) = \frac{\sqrt{\alpha^2 +
\tsnr{\beta^2}} - \alpha}{\tsnr} \label{eq: cmmse for Ornstein-Uhlenbeck process}
\eea
\item Smoothing Error
\bea
\tmmse(\tsnr) = \nu(\infty,\infty,\tsnr) = \frac{\beta^2}{2\sqrt{\alpha^2 +
\tsnr{\beta^2}}}. \label{eq: mmse for Ornstein-Uhlenbeck process}
\eea
\end{itemize}
\end{lemma}

The expressions in (\ref{eq: cmmse for Ornstein-Uhlenbeck process}) and (\ref{eq: mmse for Ornstein-Uhlenbeck process}) recover the classical expressions for the optimal causal and smoothing errors (for Gaussian inputs), due to Yovits and Jackson in \cite{YovitsJackson55}, and Wiener in \cite{Wiener42}, respectively. 

\textit{Remark:} (\ref{eq: cmmse for Ornstein-Uhlenbeck process}) and (\ref{eq: mmse for Ornstein-Uhlenbeck process}) can be easily seen to verify the following general relationship between the filtering and smoothing error for the continuous time
Gaussian channel, established in \cite{gsv2005}, i.e.
\bea
\tcmmse(\tsnr) &=& \frac{1}{\tsnr}\int_0^{\tsnr}\tmmse(\gamma)\,d\gamma. \label{eq: relation between cmmse and mmse}
\eea

\subsection{Estimation with Lookahead}
Having introduced the Ornstein-Ulhenbeck process and its properties, we now study the behavior of the MMSE with finite lookahead. \\
\textbf{Note:} For the rest of this paper we will set $\beta = 1$ in (\ref{eq: Ornstein-Uhlenbeck process evolution}), as it only involves an effective scaling of the channel SNR parameter. 

Recall that,
\begin{align} \label{def: lmmse(d,gamma)}
\lmmse(d,\gamma) &= \tvar(X_0|Y_{-\infty}^d) \\
&= \nu(\infty,d,\gamma), \label{eq: reference to nu}
\end{align}
where (\ref{eq: reference to nu}) follows from the definition of $\nu(\cdot,\cdot,\gamma)$ in Lemma \ref{lemma: nu(l,d,gamma) for Ornstein-Uhlenbeck}.
The following lemma explicitly characterizes the finite lookahead MMSE for the Ornstein-Ulhenbeck process with parameter $\alpha$, corrupted by the Gaussian channel at signal-to-noise ratio $\gamma$.
\begin{lemma}[MMSE with Lookahead for OU($\alpha$)] \label{lemma: estimation error with finite lookahead for Ornstein-Uhlenbeck}
\bea \label{eq: estimation error with finite lookahead}
\lmmse(d,\gamma) = \twopartdef{(1-e^{-2 d \sqrt{\alpha^2 + \gamma}})\text{mmse}(\gamma) + e^{-2 d \sqrt{\alpha^2 + \gamma}}\text{cmmse}(\gamma)}{d \geq 0}{e^{-2 \alpha |d|}\text{cmmse}(\gamma) + \frac{1}{2\alpha}(1-e^{-2\alpha |d|})}{d < 0.}
\eea
\end{lemma}
\begin{IEEEproof}
We will establish the expression for positive and negative values lookahead separately. 

For $d \geq 0$: We combine (\ref{def: lmmse(d,gamma)}) and Lemma \ref{lemma: nu(l,d,gamma) for Ornstein-Uhlenbeck} to obtain the closed form expression for $\lmmse(d,\gamma)$. In addition, from Lemma \ref{lemma: mse OU} we note that
\bea
\text{cmmse}(\gamma) &=& \frac{\sqrt{\alpha^2 + \gamma} - \alpha}{\gamma}, \label{eq: cmmse ou alpha}\\
\text{mmse}(\gamma) &=&  \frac{1}{2\sqrt{\alpha^2 + \gamma}}. \label{eq: mmse ou alpha}
\eea
Using the above expressions, we can express the estimation error with finite lookahead $d \geq 0$ in the following succinct manner
\bea
\lmmse(d,\gamma) &=& (1-e^{-2 d \sqrt{\alpha^2 + \gamma}})\text{mmse}(\gamma) +
e^{-2 d \sqrt{\alpha^2 + \gamma}}\text{cmmse}(\gamma).
\eea
For $d < 0$: We denote the absolute value of $d$ as $|d|$, and note that
\bea
Y_{-\infty}^{-|d|}\--X_{-|d|}\--{X_0}
\eea
forms a Markov triplet. We further note that ($X_0,X_{-|d|},Y_{-\infty}^{-|d|}$) are jointly Gaussian. In particular, we know that
\begin{align}
X_0|X_{-|d|} &\sim \Nscr \bigg(X_{-|d|} e^{-\alpha |d|},
\frac{1}{2\alpha}(1-e^{-2\alpha |d|})\bigg), \\
\tvar(X_{-|d|}|Y_{-\infty}^{-|d|}) &= \text{cmmse}(\gamma) 
\end{align}
From the above relations, it is easy to arrive at the required quantity $\tvar(X_0|Y_{-\infty}^{-|d|})$ for $d<0$,
\bea
\lmmse(d,\gamma) &=& e^{-2 \alpha |d|}\text{cmmse}(\gamma) +
\frac{1}{2\alpha}(1-e^{-2\alpha |d|}).
\eea
This completes the proof. 
\end{IEEEproof}
\begin{figure}
\begin{center}
\includegraphics[height=4in,width=6in]{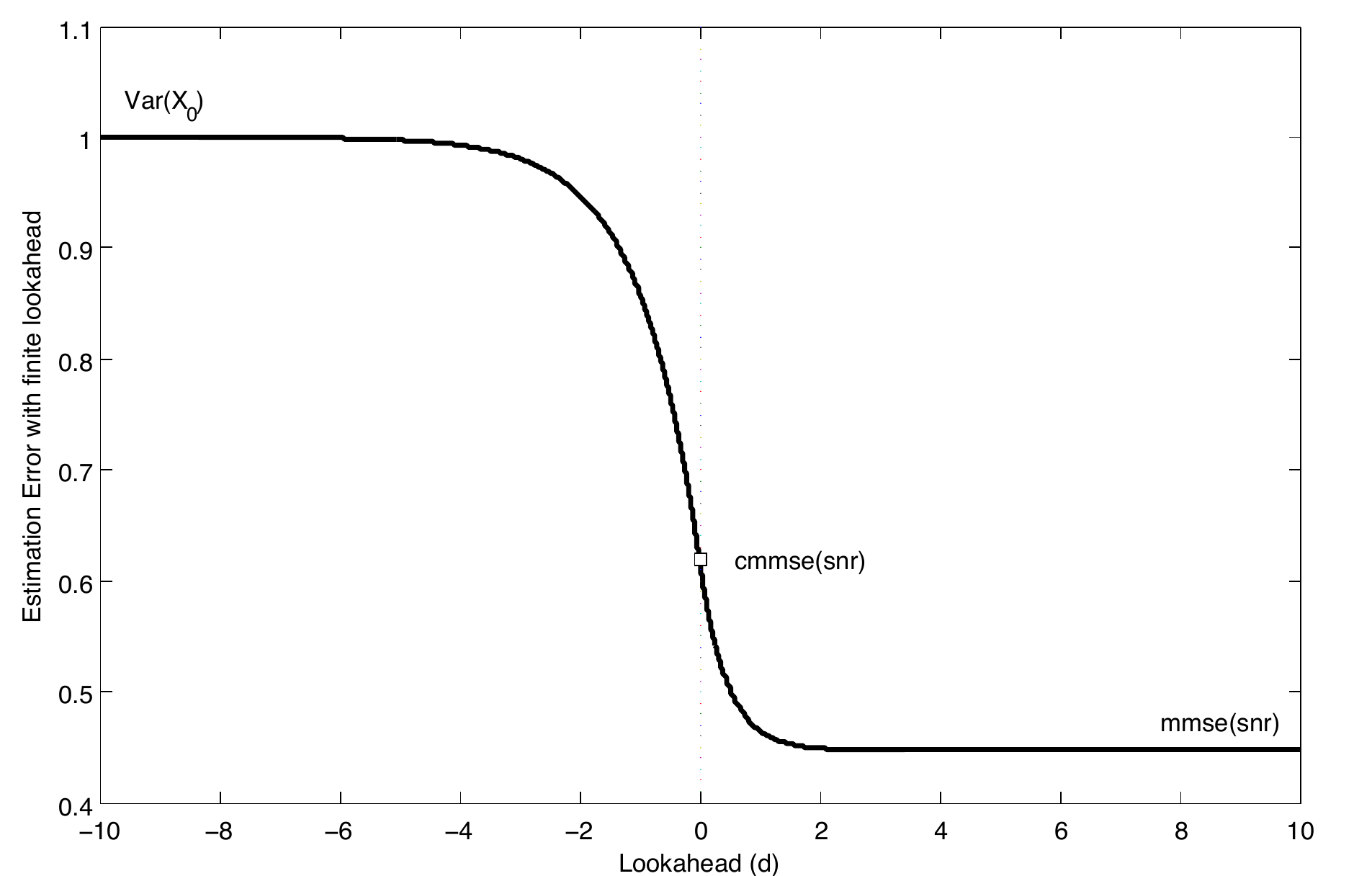}
\caption{$\lmmse(d,\tsnr)$ vs. lookahead $d$ for the Ornstein-Uhlenbeck process for $\alpha=0.5$, and $\tsnr = 1$.}
\label{fig: plot of error vs lookahead for Ornstein-Uhlenbeck process}
\end{center}
\end{figure}
A plot of the estimation error with finite lookahead for the Ornstein-Uhlenbeck process is shown in Fig. \ref{fig: plot of error vs lookahead for Ornstein-Uhlenbeck process}. It is  seen from the expression in Lemma \ref{lemma: estimation error with finite lookahead for Ornstein-Uhlenbeck}, that the asymptotes at $d = -\infty$ and $d = +\infty$ correspond to the values $R_X(0)$ and $\tmmse(\tsnr)$, respectively, for a fixed SNR level $\tsnr$.

We now focus on the case where $d \geq 0$, which corresponds to estimation with finite delay. We define, for $d \geq 0$,
\bea
p_d \doteq \frac{\lmmse(d,\gamma) - \tmmse(\gamma)}{\tcmmse(\gamma) - \tmmse(\gamma)}. \label{eq: definition of pd}
\eea
From Lemma \ref{lemma: estimation error with finite lookahead for Ornstein-Uhlenbeck}, we observe that for the Ornstein-Ulhenbeck process,
\bea
p_d &=& e^{-2 d \sqrt{\alpha^2 + \gamma}} \\
&=& e^{-\frac{d}{\text{mmse}{(\gamma)}}}, \label{eq: exponent of decay for ou}
\eea 
where (\ref{eq: exponent of decay for ou}) follows from (\ref{eq: mmse ou alpha}).  
In other words, the causal error approaches the non-causal error exponentially fast with increasing lookahead. This conveys the importance of lookahead in signal estimation with finite delay. We can state this observation as follows:

\begin{observation}
For any AWGN corrupted Ornstein-Ulhenbeck process, the mean squared error with finite positive lookahead, approaches the non-causal error exponentially fast, with decay exponent given by $\frac{1}{\text{mmse}(\gamma)}$.   
\end{observation}
%It is natural to ask whether the exponential benefit of lookahead evidenced above is true for a larger class of input processes?  A discussion to this effect is provided in Section \ref{sec: stationary gaussian processes}.

%To conclude this subsection, let us consider allowing $D$ to be the random variable denoting the lookahead in estimation. Specifically, consider the case $D \sim
%Exp(\lambda)$. We obtain the following simple expression for the expected value of mean square loss with lookahead $D$
%\bea
%\E[\lmmse(D,\gamma)] = \text{mmse}(\gamma) + \big[\text{cmmse}(\gamma) -
%\text{mmse}(\gamma)\big]\bigg(\frac{\lambda}{\lambda +
%\frac{1}{\text{mmse}(\gamma)}}\bigg)
%\eea
%In particular, for $\lambda = \frac{1}{\text{mmse}(\gamma)}$, we obtain
%\bea
%\E[\lmmse(D,\gamma)] = \frac{\text{mmse}(\gamma) + \text{cmmse}(\gamma)}{2}.
%\eea
\subsection{SNR vs Lookahead tradeoff}
One way to quantify the utility of lookahead in finite delay estimation is to
compute the corresponding gain in SNR. Specifically, we compute
the required SNR level which gives the same mmse with lookahead
$d$ as the filtering error at a fixed SNR level $\tsnr > 0$.

Let $\gamma^{*}_d(\tsnr)$ be the value of `signal to noise ratio' that provides
the same mean square error as the causal filter with zero lookahead. I.e.
\bea
\lmmse(d,\gamma^{*}_d(\tsnr)) = \tcmmse(\tsnr),
\eea
whenever a solution $\gamma^{*}_d(\tsnr) > 0$ exists.

\begin{figure}
\begin{center}
\includegraphics[height=4in,width=6in]{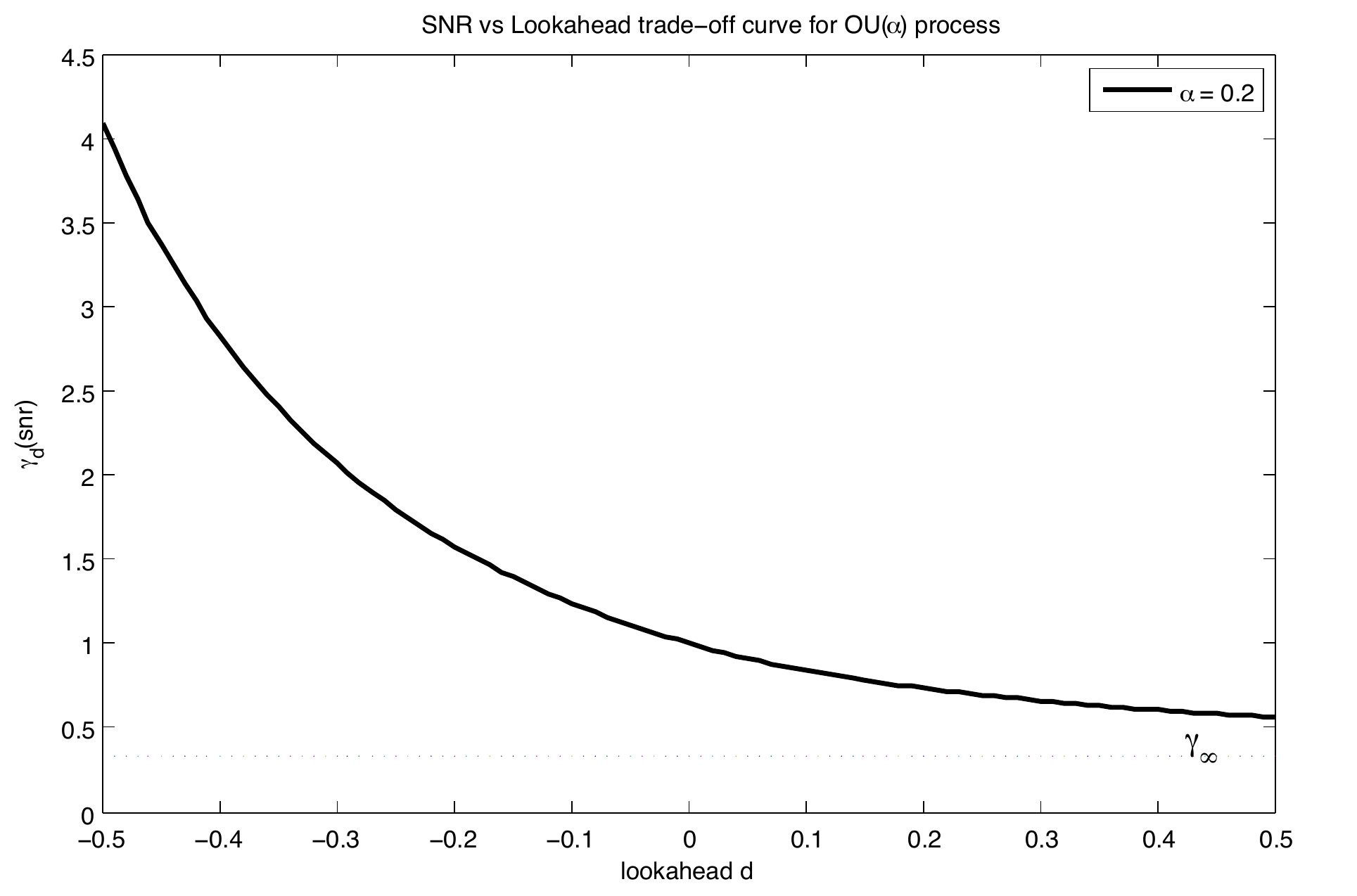}
\caption{Plot of $\gamma^{*}_d(\tsnr)$ ($\tsnr=1$) as a function of $d$ for the
Ornstein Uhlenbeck process with parameter $\alpha = 0.2$. In this case
$\gamma_\infty$ = 0.3320 and $d^{*} = -0.9935$.}
\label{fig: lookahead vs. SNR for Ornstein-Uhlenbeck process}
\end{center}
\end{figure}

We now present some general observations about $\gamma^{*}_d(\tsnr)$.
\begin{itemize}
\item $\gamma^{*}_d(\tsnr)$ is a monotonically decreasing function of $d$.
\item $\gamma^{*}_0(\tsnr) = \tsnr$.
\item Let $\gamma_{\infty}(\tsnr) = \lim_{d \to \infty} \gamma^{*}_d(\tsnr)$.
Then, we have $\tmmse(\gamma_\infty(\tsnr)) = \tcmmse(\tsnr)$. For the
Ornstein-Uhlenbeck process parameterized by $\alpha$, we get the following
closed form expression for $\gamma_\infty(\tsnr)$:
\bea
\gamma_{\infty}(\tsnr) &=& \frac{1}{(2\tcmmse(\tsnr))^2} - \alpha^2.
\eea
\item $\gamma^{*}_d(\tsnr)$ has a vertical asymptote at $d = d^{*}_{\tsnr} <
0$, where $d^{*}_{\tsnr}$ is defined as the solution to the equation
\tvar($X_0|X_{d^{*}_{\tsnr}}$) = $\tcmmse(\tsnr)$. Then we have,
\bea
\lim_{d \to d^{*}_{snr}}\gamma^{*}_d(\tsnr) = \infty
\eea
\end{itemize}
In Fig. \ref{fig: lookahead vs. SNR for Ornstein-Uhlenbeck process}, we illustrate the behavior of $\gamma_d(\tsnr)$ as a
function of lookahead, for the Ornstein-Uhlenbeck process corrupted by Gaussian
noise according to the channel in (\ref{eq: channel model}).

\section{A mixture of Ornstein-Uhlenbeck processes} \label{sec: mixture of
Ornstein-Uhlenbeck processes}
Having presented the finite lookahead MMSE for the Ornstein-Ulhenbeck process in Lemma \ref{lemma: estimation error with finite lookahead for Ornstein-Uhlenbeck}, in this section we obtain the MMSE with lookahead for the class of stochastic processes that are mixtures of Ornstein-Ulhenbeck processes. We then proceed to establish a general lower bound on  the MMSE with lookahead for the class Gaussian processes whose spectra can be decomposed as a mixture of spectra of Ornstein-Ulhenbeck processes. For the same class of Gaussian processes, we also present an upper bound for the finite lookahead MMSE, in terms of a mismatched estimation loss. 

Let $\mu(\alpha)$ be a probability measure defined on $[0, \infty)$. Let $P^{(\alpha)}$ be the law of the Ornstein-Uhlenbeck process with parameter $\alpha$. Note that each $P^{(\alpha)}$ is the law of a stationary ergodic stochastic process. We define the stationary distribution generated by taking a $\mu$-mixture of these processes:
\bea
P = \int P^{(\alpha)} \,d\mu(\alpha). \label{eq: mixture of Ornstein-Uhlenbeck}
\eea
Note that $P$ need not be Gaussian in general. 
\begin{lemma} \label{lemma: mixture of OU}
Let $X$ be a stationary stochastic process governed by law $P$ which is a $\mu$ mixture of Ornstein-Ulhenbeck processes, and is corrupted by AWGN at SNR $\gamma$. The MMSE with finite lookahead $d$ is given by
\bea
\lmmse_{P}(d,\gamma) = \int \lmmse_{\alpha} (d,\gamma)
\,d\mu({\alpha}), \label{eq: lmmse for Ornstein-Uhlenbeck mixture}
\eea
where $\lmmse_{\alpha} (d,\gamma)$ is (as defined in Lemma \ref{lemma: estimation error with finite lookahead for Ornstein-Uhlenbeck}) the corresponding quantity for estimating an Ornstein-Ulhenbeck process with parameter $\alpha$. 
\end{lemma}
The proof of Lemma \ref{lemma: mixture of OU} follows in one line, upon observing that the underlying `active mode' is eventually precisely learned from the infinitely long observation of the noisy process.  This relation allows us to compute the minimum mean squared error with finite lookahead for the large
class of processes that can be expressed as mixtures of Ornstein-Uhlenbeck processes.

As another important corollary of this discussion, consider any Gaussian process $G$ whose spectrum $S_G$ can be expressed as a mixture of
spectra of Ornstein-Uhlenbeck processes, for some appropriate mixing measure $\mu(\alpha)$,i.e.
\bea
S_G = \int S_{\alpha} \, d \mu (\alpha), \label{eq: gaussian process with mixture spectrum}
\eea
where $S_{\alpha}$ denotes the spectrum of the Ornstein-Ulhenbeck process with parameter $\alpha$. The approach outlined above provides us with a computable lower bound on the minimum mean squared error with fixed lookahead $d$ (under AWGN) for the process $G$, which we state in the following Lemma.
\begin{lemma} \label{lemma: lower bound on lmmse}
For a Gaussian process $G$ with spectrum as in (\ref{eq: gaussian process with mixture spectrum}), the finite lookahead mmse has the following lower bound:
\bea
\lmmse_{G}(d,\gamma) \geq \int \lmmse_{\alpha}(d,\gamma) \,d\mu(\alpha), \label{eq: lower bound on lmmse}
\eea
where $\lmmse_{\alpha}(d, \gamma)$ is characterized explicitly in Lemma \ref{lemma: estimation error with finite lookahead for Ornstein-Uhlenbeck}. 
\end{lemma}
To see why (\ref{eq: lower bound on lmmse}) holds note that its right hand side represents the mmse at lookahead $d$ associated with the process whose law is expressed in (\ref{eq: mixture of Ornstein-Uhlenbeck}), while the left side corresponds to this mmse under a Gaussian source with the same spectrum.

In the following example, we will illustrate the bound in (\ref{eq: lower bound on lmmse}). In Section \ref{sec: gaussian processes}, we discuss the computation of the mmse with lookahead for any stationary Gaussian process, of a known spectrum. This computation, based on Wiener-Kolmogorov theory, relies on the spectral factorization of the spectrum of the AWGN corrupted process. This factorization is, in general, difficult to perform. Thus, a bound on the mmse with lookahead, such as in (\ref{eq: lower bound on lmmse}), for a Gaussian process whose spectrum can be expressed as a mixture of spectra of Ornstein-Ulhenbeck processes, is quite useful. 

A natural question that emerges from the above discussion is, what functions can be decomposed into mixtures of spectra of Ornstein-Ulhenbeck processes ? To answer this question, note that one can arrive at any spectrum $S_G$ which can be expressed (upto a multiplicative constant) as 
\bea
S_G(\omega) = \int_0^\infty \frac{1}{\alpha^2 + \omega^2} \, d\mu(\alpha) \label{eq: spectrum decomposition}
\eea
where we use (\ref{eq: PSD of Ornstein-Uhlenbeck process}) to characterize $S_\alpha (\omega)$. Equivalently, in the time domain, the desired auto-correlation function is expressible as 
\bea
R_G(\tau) = \int_0^\infty e^{-\alpha |\tau|}\, \frac{d\mu(\alpha)}{\alpha},   \label{eq: autocorrelation decomposition}
\eea
which can be viewed as a real-exponential transform of the function $\mu$. However, as can be seen from (\ref{eq: spectrum decomposition}), the spectrum $S_G(\omega)$ is always constrained to be monotonically decreasing with $\omega$. This shows that the space of functions that can be candidates for the spectrum of the process G, is not exhausted by the class of functions decomposable as spectra of Ornstein-Ulhenbeck processes. 
\subsection{Illustration of bound in Lemma \ref{lemma: lower bound on lmmse} } \label{example: lower bound}
\begin{figure}
\begin{center}
\includegraphics[height=4in,width=6in]{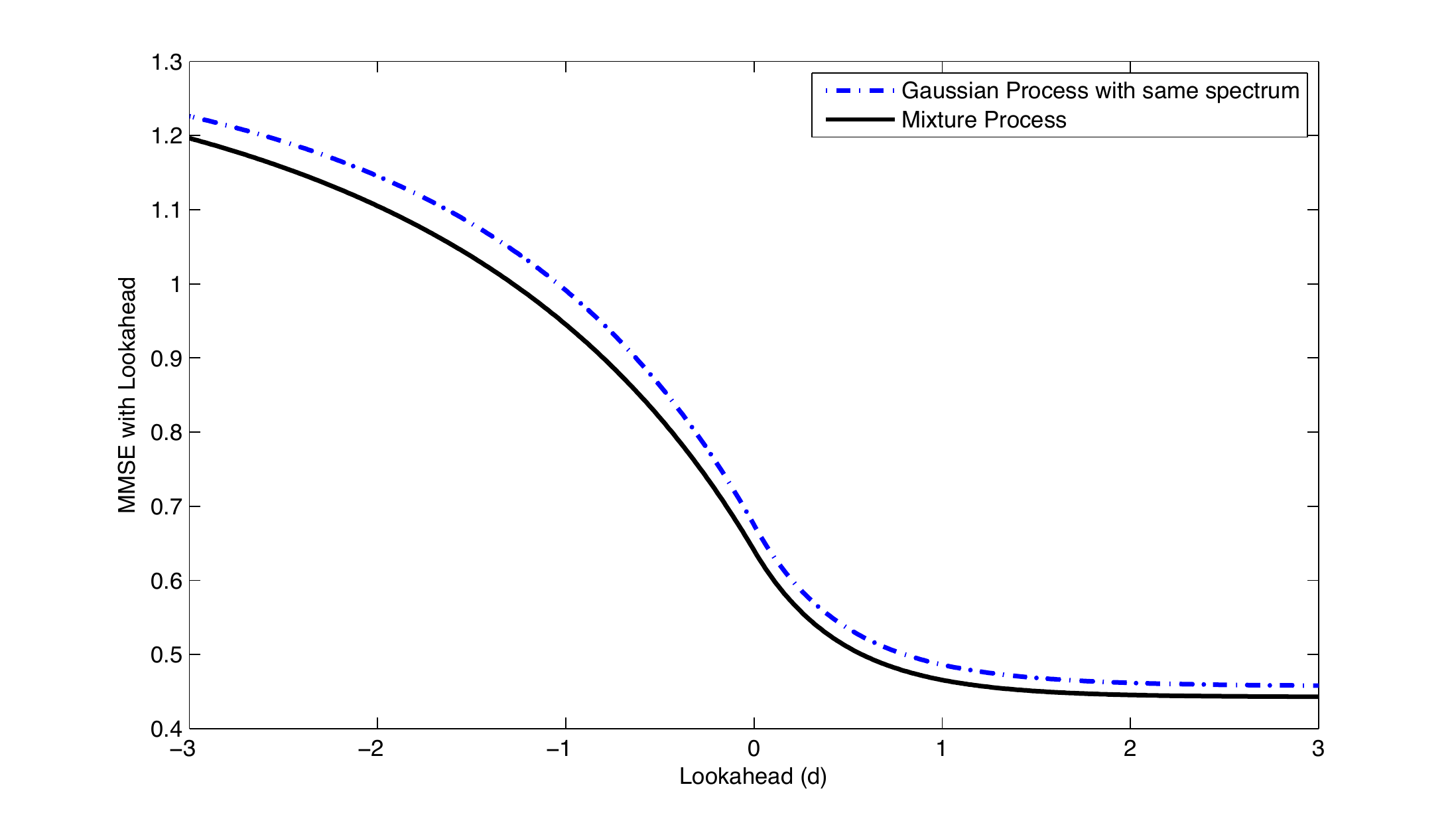}
\caption{Illustration of the bound in (\ref{eq: lower bound on lmmse}). Plot of MMSE with lookahead for the processes $\mathbf{G}$ and $\mathbf{X}$ in Example \ref{example: lower bound}, for $\tsnr=1$. $\mathbf{X}$ is a mixture of two Ornstein-Ulhenbeck processes with parameters $\alpha_1 = 0.75$ and $\alpha_2 = 0.25$ respectively. $\mathbf{G}$ is a stationary Gaussian process with the same spectrum as $\mathbf{X}$. }
\label{fig: illustration of bound}
\end{center}
\end{figure}

We consider the simple scenario of a process which is the equal mixture of two Ornstein-Ulhenbeck processes, $\mathbf{X}_1$ and $\mathbf{X}_2$, parametrized by $\alpha_1$ and $\alpha_2$ respectively. Specifically, let us define a stationary continous-time process $\mathbf{X}$ as
\begin{align}
\mathbf{X} = \twopartdef{\mathbf{X}_1}{\text{w.p.} \frac{1}{2}}{\mathbf{X}_2}{\text{w.p.} \frac{1}{2}}  
\end{align}
Note that the process $\mathbf{X}$ will not be Gaussian in general. The spectrum $S_{\mathbf{X}}(\cdot)$ of the mixture, will be given by the mixture of the spectra of the constituent Ornstein-Ulhenbeck processes. I.e.
\begin{align} 
S_{\mathbf{X}}(\cdot) =  \frac{1}{2}S_{\mathbf{X}_1}(\cdot) + \frac{1}{2}S_{\mathbf{X}_2}(\cdot),
\end{align}
where $S_{\mathbf{X}_i}$, denotes the spectrum of $\mathbf{X}_i$, $i = 1,2$. Now consider a stationary Gaussian process $\mathbf{G}$ with the same spectrum, i.e.
\begin{align}
S_\mathbf{G}(\cdot) = S_{\mathbf{X}}(\cdot).
\end{align} 
We will consider the noise corruption mechanism in (\ref{eq: channel model}) at a signal-to-noise ratio $\tsnr$. Note that Lemma \ref{lemma: estimation error with finite lookahead for Ornstein-Uhlenbeck}, in conjunction with (\ref{eq: lmmse for Ornstein-Uhlenbeck mixture}), allow us to compute the mmse with lookahead $\lmmse_{\mathbf{X}}(d,\tsnr)$, for the $\mathbf{X}$ process:
\begin{align}
\lmmse_{\mathbf{X}}(d,\tsnr) = \frac{1}{2} \lmmse_{\mathbf{X}_1}(d,\tsnr) + \frac{1}{2} \lmmse_{\mathbf{X}_2}(d,\tsnr),
\end{align}  
for all $d$. Note that when $d = -\infty$, the quantity $\lmmse(-\infty, \tsnr)$ will be given by the stationary variance of the underlying process. Since this quantity depends only on the spectrum of the stationary process, the lower bound in (\ref{eq: lower bound on lmmse}) will coincide with the estimation error for the $\mathbf{G}$ process at $d = -\infty$. For the $\mathbf{G}$ process, we can analytically compute the mmse with finite lookahead (\ref{eq: lmmse of mixture example}) as well, using the method outlined in Section \ref{sec: gaussian processes}. The reader is referred to Appendix \ref{appendix: lmmse for gaussian spectrum} for a detailed discussion on computing the estimation error with finite lookahead for the $\mathbf{G}$ process under any $\alpha_1, \alpha_2$. For the purposes of this example, we illustrate the bound (\ref{eq: lower bound on lmmse}) for the case when $\alpha_1 = 0.75$ and $\alpha_2 = 0.25$. In Fig. \ref{fig: illustration of bound}, we present the mmse with lookahead for process $\mathbf{G}$, as well as the lower bound, given by the corresponding quantity for $\mathbf{X}$. For the given choice of parameters, at $\tsnr=1$: $\lmmse_{\mathbf{X}}(-\infty,\tsnr)=\lmmse_{\mathbf{G}}(-\infty,\tsnr)= 1.333$; $\lmmse_{\mathbf{X}}(+\infty,\tsnr) = 0.4425$; $\lmmse_{\mathbf{G}}(+\infty,\tsnr)=0.4568$.

\subsection{Towards deriving an Upper Bound}
Note that Lemma \ref{lemma: lower bound on lmmse} gives us a computable lower bound for the mmse with lookahead for a Gaussian process $G$ whose spectrum can be expressed as a mixture of spectra of Ornstein-Ulhenbeck processes, under an appropriate mixture $\mu$. %: (0,\infty) \to \Re^{+} \cup \{0\}$. 

Define, the mismatched mmse with lookahead for a filter that assumes the underlying process has law $P^{(\beta)}$ [Ornstein-Ulhenbeck process with parameter $\beta$], whereas the true law of the process is $P^{(\alpha)}$
\bea
\lmmse_{\alpha,\beta} \, (d, \gamma) = \E_{\alpha} [ (X_0 - \E_{\beta}[X_0 | Y_{-\infty}^d])^2 ], \label{eq: mismatched mmse with lookahead}
\eea
where the outer expectation is with respect to the true law of the underlying process $P^{(\alpha)}$, while the inner expectation is with respect to the mismatched law $P^{(\beta)}$. 
Note that the mismatched filter which assumes that the underlying signal is an Ornstein-Ulhenbeck process with parameter $\beta$, is in particular, a linear filter. The process $G$ is Gaussian, and hence the optimal filter for $G$ is also linear. Thus, for any fixed $\beta$, the mismatched filter thus defined, will be suboptimal for process $G$. This yields the following natural upper bound, which in conjunction with Lemma \ref{lemma: lower bound on lmmse} can be stated as:
\begin{lemma}[Upper and lower bound in terms of mixture of Ornstein-Ulhenbeck spectra] \label{lemma: upper and lower bound}
\bea
\int \lmmse_{\alpha}(d,\gamma) \,d\mu(\alpha) \leq \lmmse_{G}(d,\gamma) \leq \min_{\beta > 0}\int \lmmse_{\alpha, \beta}\,(d,\gamma) \,d\mu(\alpha).
\eea
\end{lemma}
We should note that the upper bound stated in Lemma \ref{lemma: upper and lower bound} can be hard to compute for continuous time processes, even though the filter applied is linear.  In summary, analysis of $\lmmse(\cdot,\gamma)$ for the Ornstein-Uhlenbeck process provides us with a rich class of processes for which we can compute the finite lookahead estimation error. And for a class of Gaussian processes (\ref{eq: autocorrelation decomposition}), we have a lower and upper bound on this quantity. These characterizations may be used to approximate the behavior of the MMSE with finite lookahead for a large class of Gaussian processes, instead of obtaining the optimal solution which relies on spectral factorization, and can be cumbersome. 

\section{Stationary Gaussian Processes} \label{sec: stationary gaussian processes}
In this section, we focus exclusively on Gaussian processes having a known spectrum. We are interested in obtaining the MMSE with finite lookahead in estimating such a process when corrupted by AWGN at a fixed SNR level $\gamma$. 
\subsection{Calculation of MMSE with finite lookahead via Spectral Factorization} \label{sec: gaussian processes}
Let $\{X_t, t \in \Re\}$ stationary and Gaussian, with power spectral density $S_X(\omega)$, be the input to the continuous-time Gaussian channel at SNR $\gamma$. $Y_{(\cdot)}$, the noise corrupted version of $X_{(\cdot)}$, is stationary, Gaussian and has PSD $S_Y(\omega) =  1+ \gamma S_X(\omega)$. Let $S_Y^+(\omega)$ be the Wiener-Hopf factorization of $S_Y(\omega)$, i.e. a function satisfying $\left|S_Y^+(\omega)\right|^2=S_Y(\omega)$ with $1/S_Y^+(\omega)$ being the transfer function of a stable causal filter. This factorization exists whenever the Paley-Wiener conditions are satisfied for $S_Y(\omega)$. Denote by $\Ytil_t$ the output of the filter $1/S_Y^+(\omega)$ applied on $Y_t$, then $\Ytil_t$ is a standard Brownian motion. This theory is well developed and the reader is encouraged to peruse any standard reference in continuous-time estimation for details (cf., e.g., \cite{Kailath00} and references therein ).

Let $h(t)$ be the impulse response of the filter with transfer function
\bea
H(\omega)=\frac{\sqrt{\gamma}S_{X}(\omega)}{S_Y^{-}(\omega)}, \label{eq: def wiener-kolmogorov H(w)}
\eea
with $S_Y^{-}(\omega)=S_Y^{+}(-\omega)$. The classical result by Wiener and Hopf \cite{WienerHopf1931}, applied to non-causal filtering of stationary
Gaussian signals can be stated as,
\bea
\E\left[X_{0}|Y_{-\infty}^{\infty}\right]=\E\left[X_{0}|\tilde{Y}_{-\infty}^{\infty}\right]=\int_{-\infty}^{\infty}h(-t)d\tilde{Y}_{t}.
\eea
Moreover, an expression for the finite-lookahead MMSE estimator of
$X_{0}$ can be immediately derived, using the fact that $\Ytil_t$
is both a standard Brownian motion and a reversible causal function of the
observations:
\bea
\E\left[X_{0}|Y_{-\infty}^{d}\right] &=& \E\left[X_{0}|\tilde{Y}_{-\infty}^{d}\right]=\E\left[\E\left\{ X_{0}|\tilde{Y}_{-\infty}^{\infty}\right\} |\tilde{Y}_{-\infty}^{d}\right]\\
 &=& \E\left[\int_{-\infty}^{\infty}h(-t)d\tilde{Y}_{t}|\tilde{Y}_{-\infty}^{d}\right]=\int_{-\infty}^{d}h(-t)d\tilde{Y}_{t}.
\eea

Using again the fact that $\Ytil_{t}$ is a brownian motion, as well
as the orthogonality property of MMSE estimators, we can obtain a simple
expression for $\textrm{lmmse}(d,\gamma)$:
\bea
\lmmse(d,\gamma) &=& \E\left(X_{0}-\E\left[X_{0}|Y_{-\infty}^{d}\right]\right)^{2}=\E\left(X_{0}-\E\left[X_{0}|\Ytil_{-\infty}^{d}\right]\right)^{2}\\
 &=& \E\left(X_{0}-\E\left[X_{0}|\Ytil_{-\infty}^{\infty}\right]\right)^{2}+\E\left(\E\left[X_{0}|\Ytil_{-\infty}^{\infty}\right]-\E\left[X_{0}|\Ytil_{-\infty}^{d}\right]\right)^{2}\\
 &=& \tmmse(\gamma)+\E\left(\int_{d}^{\infty}h(-t)d\tilde{Y}_{t}\right)^{2}=\tmmse(\gamma)+\int_{-\infty}^{-d}h^{2}(t)dt. \label{eq: lmmse in terms of estimation filter impulse response}
\eea
This classical formulation of the lookahead problem for stationary Gaussian processes shows us that the MMSE with lookahead behaves gracefully with the lookahead $d$, and is intimately connected to the impulse response $h(\cdot)$ of the filter induced by the Wiener spectral factorization process. In particular, that the solution to the lookahead problem for each value of $d$, can be approached in this single unified manner as shown in (\ref{eq: lmmse in terms of estimation filter impulse response}), is quite satisfying.   
%From this expression it is simple to see that $\lmmse(d,\gamma)$ approaches $\tmmse(\gamma)$ exponentially fast as $d\tends\infty$ iff the negative tail of $h(t)$ decays exponentially fast.

\subsection{Processes with a rational spectrum}
Let $S_X(\omega)$ be a rational function, i.e. of the form $\frac{P(\omega)}{Q(\omega)}$ with $P,Q$ being finite order polynomials in $\omega$. In this case $S_Y(\omega)$ is also a rational function, and the Wiener-Hopf factorization can be preformed simply by factorizing the numerator and denominator of $S_Y(\omega)$ and composing $S^+_Y(\omega)$ only from the stable zeros and poles.

Recall the definition of $p_d$ in (\ref{eq: definition of pd}),
\bea
p_d \doteq \frac{\lmmse(d,\gamma) - \tmmse(\gamma)}{\tcmmse(\gamma) - \tmmse(\gamma)}. 
\eea
Clearly, $p_0=1$ and $\lim_{d\tends\infty}p_d=0$. The value $p_d$ goes to zero in the same rate as $\lmmse$ converges to $\tmmse$. For the Ornstein-Ulhenbeck process we observed that $p_d$ converges exponentially to zero. In the following lemma, we generalize this result to include all Gaussian input processes with rational spectra

\begin{lemma} \label{lemma: rational spectra}
For any Gaussian input process with a rational spectrum, $\lmmse(d,\tsnr)$ approaches $\tmmse(\tsnr)$ exponentially fast as $d\tends\infty$. Equivalently, $p_d \to 0$ exponentially, as $d\to \infty$.
\end{lemma}
\begin{IEEEproof}Since $S^+_Y(\omega)$ is a rational function, so is $H(\omega)$ defined in (\ref{eq: def wiener-kolmogorov H(w)}). Therefore, $h(t)$ must be a finite sum of exponentially decreasing functions [should use a ref here] and consequently $\int_{-\infty}^{-d}h^{2}(t)dt$ must also decrease exponentially as $d\tends\infty$. This in conjunction with the relation in (\ref{eq: lmmse in terms of estimation filter impulse response}) concludes the proof.
\end{IEEEproof}

As an illustration, we consider an equal mixture of two Ornstein-Ulhenbeck processes with parameters $\alpha_1$ and $\alpha_2$, as in the example of the previous section. In Appendix \ref{appendix: lmmse for gaussian spectrum} the mmse with lookahead is explicitly computed for a Gaussian process with this spectrum, acting as input to the channel. From these calculations, one can observe that the corresponding expression for $p_d$ in (\ref{eq: pd for mixture example}) vanishes exponentially with increasing lookahead.  However, it is important to note that there exist input spectra for which the decay of $p_d$ is not exponentially fast. 

\begin{figure}
\begin{center}
\includegraphics[height=4in,width=6in]{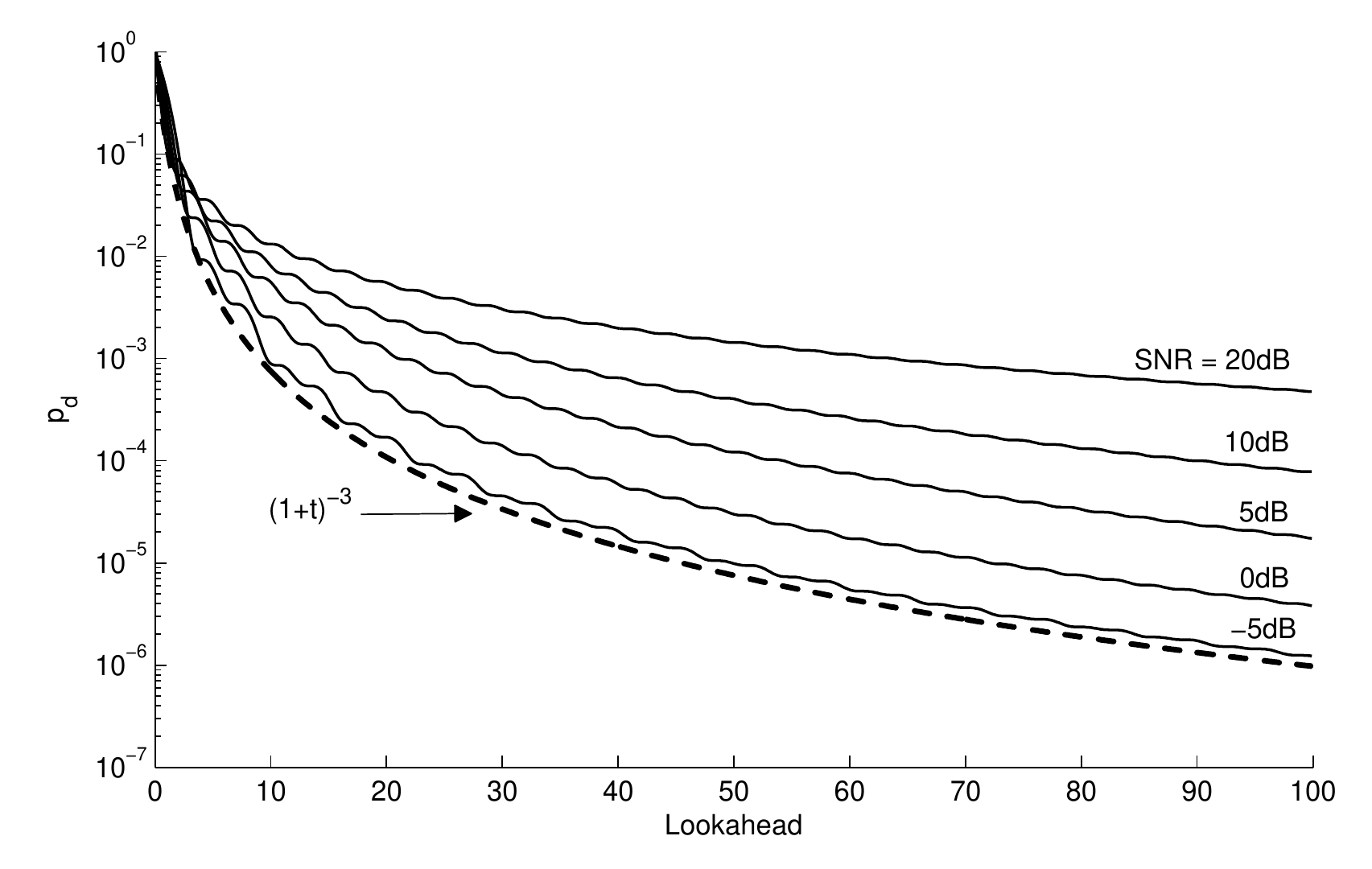}
\caption{Plot of $p_d$ as defined in (\ref{eq: definition of pd}), vs. lookahead in various SNRs, for input power spectrum $\left(1-|\omega|\right)\mathbf{1}_{\left\{ \left|\omega\right|\leq1\right\} }$. The finite lookahead MMSE is seen to converge to the non-causal MMSE at an approximately cubic rate.}
\label{fig: lmmse of bandlimited process}
\end{center}
\end{figure}

As an example, consider the input power spectrum
\bea
S^{\text{triang}}_X(\omega)=\left(1-|\omega|\right)\mathbf{1}_{\left\{ \left|\omega\right|\leq1\right\} },
\eea
and 
In Figure \ref{fig: lmmse of bandlimited process} we plot the behaviour of $p_d$ as a function of $d$ for input power spectrum $S^{\text{triang}}_X(\omega)$ and different SNR's. This demonstrates the polynomial rate of convergence of the finite lookahead estimation error towards to non-causal MMSE. The values of $\lmmse(d,\tsnr)$ were found by numerically approximating $h(t)$, using FFT in order to perform the factorization of $S_Y(\omega)$.

%We remark that band-limitedness is a sufficient but not a necessary condition for polynomial behaviour of the finite-lookahead estimation error. For example, numerical experiments suggest that the input spectrum $S_X(\omega)=e^{-|\omega|}$ yields polynomial convergence as well.
%%[viewport=162 267 642 528,clip,scale=0.85]

\section{Information Utility Of Lookahead} \label{sec: information utility}
Consider a stationary process $X_t$ observed through the
continuous-time Gaussian channel (\ref{eq: channel model}) at SNR level
$\gamma$. In the previous sections we have tried to ascertain the benefit of
lookahead in mean squared estimation. We now address the question, how much information does lookahead provide in general, and whether this quantity
has any interesting connections with classical estimation theoretic objects.

For $\tau>0$, let us define the Information Utility $U(\cdot)$ as a function of lookahead $\tau$, to be
\bea
U(\tau) = I(X_0;Y_0^\tau|Y_{-\infty}^0).  \label{eq: define U}
\eea

When the input process is Gaussian,
\bea
U(\tau) &=& h(X_0|Y_{-\infty}^0)-h(X_0|Y_{-\infty}^\tau)=
\frac{1}{2}\log\left(2{\pi}e\tvar(X_0|Y_{-\infty}^0)\right)-
\frac{1}{2}\log\left(2{\pi}e\tvar(X_0|Y_{-\infty}^\tau)\right)\\
&=& \frac{1}{2}\log\left(\frac{\tcmmse(\tsnr)}{\lmmse(\tau,\tsnr)}\right).
\eea
Rearranging the terms, we have
\bea
\lmmse(\tau,\tsnr) = \tcmmse(\tsnr)\exp({-2U(\tau)}).\label{eq: Ut lmmse gaussian}
\eea

Furthermore, when the input is non-Gaussian but $h(X_0|Y_{-\infty}^\tau)$ is well-defined for every $\tau \geq 0$,
\bea
h(X_0|Y_{-\infty}^\tau)\leq\E_{y_{-\infty}^\tau}\left[\frac{1}{2}\log\left(2{\pi}e\tvar(X_0|Y_{-\infty}^\tau=y_{-\infty}^\tau)\right)\right]
\leq\frac{1}{2}\log\left(2{\pi}e\lmmse(\tau,\tsnr)\right).
\eea
The first inequality is due to the fact that the Gaussian distribution has maximum entropy under a variance constraint, and the second inequality follows from Jensen's inequality.
Rearranging the terms once more, we have that for every stationary input process,
\bea
\lmmse(\tau,\tsnr) {\geq} N(X_0|Y_{-\infty}^0)\exp ({-2U(\tau)}),
\eea
where $N(Z)=\frac{1}{2{\pi}e}\exp ({2h(Z)})$ is the entropy power functional.

We now present a relation between $U(\cdot)$ and the MMSE via a differential equation. Consider an infinitesimal increment $d\tau$ in $U(\tau)$,
\bea
U(\tau+d\tau)-U(\tau)=I(X_0;Y_\tau^{\tau+d\tau}|Y_{-\infty}^\tau)=I(X_\tau^{\tau+d\tau};Y_\tau^{\tau+d\tau}|Y_{-\infty}^
\tau)-I(X_\tau^{\tau+d\tau};Y_\tau^{\tau+d\tau}|Y_{-\infty}^\tau,X_0).
\eea
The last equality is due to the Markov Chain relationship $X_0-(X_\tau^{\tau+\,d\tau},Y_{-\infty}^\tau)-Y_\tau^{\tau+d\tau}$. Using the time-incremental channel argument introduced in \cite[Section III.D]{gsv2005}, we are able to write,
\begin{align}
I(X_\tau^{\tau+d\tau};Y_\tau^{\tau+d\tau}|Y_{-\infty}^
\tau) = \frac{1}{2}\,d\tau\cdot{\tsnr}{\cdot}\tvar(X_\tau|Y_{-\infty}^\tau) + o(d\tau),
\end{align}
and
\begin{align}
I(X_\tau^{\tau+d\tau};Y_\tau^{\tau+d\tau}|Y_{-\infty}^\tau,X_0) = \frac{1}{2}\,d\tau{\cdot}{\tsnr}{\cdot}\tvar(X_\tau|Y_{-\infty}^\tau,X_0) + o(d\tau).
\end{align}
The expression for the time derivative of the information utility function is therefore,
\bea
U'(\tau)=\frac{\tsnr}{2}\left(\tvar(X_\tau|Y_{-\infty}^\tau)-\tvar(X_\tau|Y_{-\infty}^\tau,X_0)\right).\label{eq: derivative of Ut}
\eea
Since the input is assumed stationary, $\tvar(X_\tau|Y_{-\infty}^\tau)=\tvar(X_0|Y_{-\infty}^0)=\tcmmse(\tsnr)$. We notice that $\tvar(X_\tau|Y_{-\infty}^\tau,X_0)$ is the causal MMSE in estimating $X_\tau$ when $X_0$ is known. The value of $U'(\tau)$ is therefore intimately connected to the effect of initial conditions on the causal MMSE estimation of the process. In particular,
\bea
U'(0)=\frac{\tsnr}{2}\tcmmse(\tsnr),
\eea
meaning that the added value in information of an infinitesimal lookahead is proportional to the causal MMSE.

Noticing that $U(0)=0$, we may integrate (\ref{eq: derivative of Ut}) to obtain,
\bea
U(\tau)=\frac{\tsnr}{2}\left[\tau\,\tcmmse(\tsnr)-\int_0^\tau \tvar(X_s|Y_{-\infty}^s,X_0)\,ds\right].\label{eq: Ut MMSE stationary}
\eea

Specializing to stationary Gaussian input processes, and joining equations (\ref{eq: Ut lmmse gaussian}) and (\ref{eq: Ut MMSE stationary}), we obtain
\bea
\lmmse(\tau,\tsnr)=\tcmmse(\tsnr)\exp{\left[-\tsnr\left(\tau \tcmmse(\tsnr)-\int_0^\tau \tvar(X_s|Y_{-\infty}^s,X_0)\,ds\right)\right]}.\label{eq: lmmse from Ut MMSE}
%\left{-\tsnr\left[t\tcmmse(\tsnr)-\int_0^tdsVar(X_s|Y_{-\infty}^s,X_0)\right]\right}.
\eea

For the Ornstein-Uhlenbeck process, $\tvar(X_s|Y_{-\infty}^s,X_0)$ can be calculated explicitly. This, in conjunction with (\ref{eq: lmmse from Ut MMSE}), provides an alternative derivation of the finite lookahead MMSE in the Ornstein-Uhlenbeck process, that is based on Information-Estimation arguments. The reader is referred to Appendix \ref{appendix: OU via Ut} for more details.

\section{Generalized Observation Model} \label{sec: generalized observation
model}
In this section, we present a new observation model to understand the
behavior of estimation error with lookahead. Consider a stationary
continuous-time stochastic process $X_t$. The observation model is still additive Gaussian noise, where the SNR level of the channel has a jump at $t=0$. Letting $Y_t$ denote the channel output, we describe the
channel as given below.
\bea \label{step snr channel}
\,dY_t &=& \left\{
  \begin{array}{l l}
    \sqrt{\tsnr}\,X_t\,dt + \,dW_t & \quad \text{t $\leq$ 0}\\
    \sqrt{\gamma}\,X_t\,dt + \,dW_t & \quad \text{t $>$ 0}\\
  \end{array} \right.
\eea
where, as usual, $W_{\cdot}$ is a standard Brownian motion independent of $\mathbf{X}$. Note that for
$\gamma \neq \tsnr$, the $(X_t,Y_t)$ process is not jointly stationary. Letting
$d, l \geq 0$, we define the finite lookahead estimation error at time $d$ with
lookahead $l$ as
\bea
f(\tsnr,\gamma,d,l) &=& \tvar(X_d|Y_{-\infty}^{l+d}). \label{eq: function f}
\eea
We call this a generalized observation model, as for $\gamma = \tsnr$, we
recover the usual time-invariant Gaussian channel. For instance, we note that
the error $f$ reduces to the filtering, smoothing or finite lookahead
errors, depending on the parameters $\gamma$, $l$ and $d$ as:
\begin{align}
\tcmmse(\tsnr) &= f(\tsnr,\tsnr,d,0) \label{eq: f-cmmse
check}\\
\tmmse(\tsnr) &= f(\tsnr,\tsnr,d,\infty)  \label{eq:
f-mmse check}\\
\lmmse(l,\tsnr) &= f(\tsnr,\tsnr,t,l) \label{eq:
f-dmmse check}
\end{align}
In the following we relate the estimation error with finite lookahead for the observation model described in
(\ref{step snr channel}), with the original filtering error.
\begin{theorem} \label{thm:
relation between f and cmmse}
Let $X_t$ be any finite variance continuous time stationary process which is
corrupted by the Gaussian channel in (\ref{step snr channel}). Let $f$ be as
defined in (\ref{eq: function f}). For fixed $\tsnr > 0$ and $T >0$ let
$\Gamma$ $\sim U[0,\tsnr]$ and $L$ $\sim U[0,T]$ be independent random variables. Then
\bea
\tcmmse(\tsnr) = \E_{\Gamma,L}[f(\tsnr,\Gamma,T-L,L)] \label{eq: expectation identity}
\eea
\end{theorem}
\begin{IEEEproof}
Before proving this result, we take a detour to consider a stationary continuous-time stochastic process
$X_0^T$ which is governed according to law $P_{X_0^T}$ and is observed in the
window $t \in [0,T]$ through the continuous-time Gaussian channel (\ref{eq:
channel model}) at SNR level $\tsnr$. We define the operators which evaluate the
filtering and smoothing estimation errors for this process model, as follows:
\bea
\underline{\tcmmse}(P_{X^T},\tsnr) = \int_0^T \E[(X_t - \E[X_t|Y^t])^2]\,dt \\
\underline{\tmmse}(P_{X^T},\tsnr) = \int_0^T \E[(X_t - \E[X_t|Y^T])^2]\,dt
\eea
Further, \cite{gsv2005} gives us the celebrated relationship between the
causal and non-causal estimation errors:
\bea
\underline{\tcmmse}(P_{X^T},\tsnr) &=&
\frac{1}{\tsnr}\int_0^{\tsnr}\underline{\tmmse}(P_{X^T},\gamma)\,d\gamma.
\eea

We now return to the auxiliary observation model introduced above. We note that
the causal and non-causal estimation errors of process $X_t$ at signal to noise
ratio $\tsnr$, can be written in terms of
\bea
\tcmmse(\tsnr) &=&
\frac{1}{T}\E_{\tsnr}\bigg[\underline{\tcmmse}(P_{X_{0}^{T}|Y_{-\infty}^{0}},
\tsnr)\bigg] \\
\tmmse(\tsnr) &=&
\frac{1}{T}\E_{\tsnr}\bigg[\underline{\tmmse}(P_{X_{0}^{T}|Y_{-\infty}^{0}},
\tsnr)\bigg],
\eea
where $\E_{\tsnr}$ denotes expectation over $Y_{-\infty}^{0}$. And, now
employing the relation between the two quantities above, we get
\bea
\frac{1}{T}\E_{\tsnr}\bigg[\underline{\tcmmse}(P_{X_{0}^{T}|Y_{-\infty}^{0}},
\tsnr)\bigg]  =
\frac{1}{T}\E_{\tsnr}\bigg[\frac{1}{\tsnr}\int_0^{\tsnr}\underline{\tmmse}(P_{X_
{0}^{T} |Y_{-\infty}^{0}}, \gamma)\,d\gamma\bigg]
\eea
Using the definition of $f$ and inserting in above, we get the following
intergral equation which holds for every stationary process $X_t$ observed
through the continuous Gaussian channel as described in (\ref{step snr channel})
\bea
f(\tsnr,\tsnr,x,0) =
\frac{1}{T \cdot \tsnr}\int_0^{\tsnr}\int_0^{T}f(\tsnr,\gamma,t,T-t)\,dt\,d\gamma, \label{eq: identity for f}
\eea
for all T and x. Note that the left hand side of (\ref{eq: identity for f}) is nothing but the causal squared error at SNR level $\tsnr$. Note, in
particular, that for independent random variables $\Gamma$ $\sim U[0,\tsnr]$ and $L$ $\sim U[0,T]$, (\ref{eq: identity for f}) can be
expressed as
\bea
\tcmmse(\tsnr) = \E_{\Gamma,L}[f(\tsnr,\Gamma,T-L,L)].
\eea
\end{IEEEproof}
\begin{figure}
\begin{center}
\includegraphics[height=2.2in,width=3.3in]{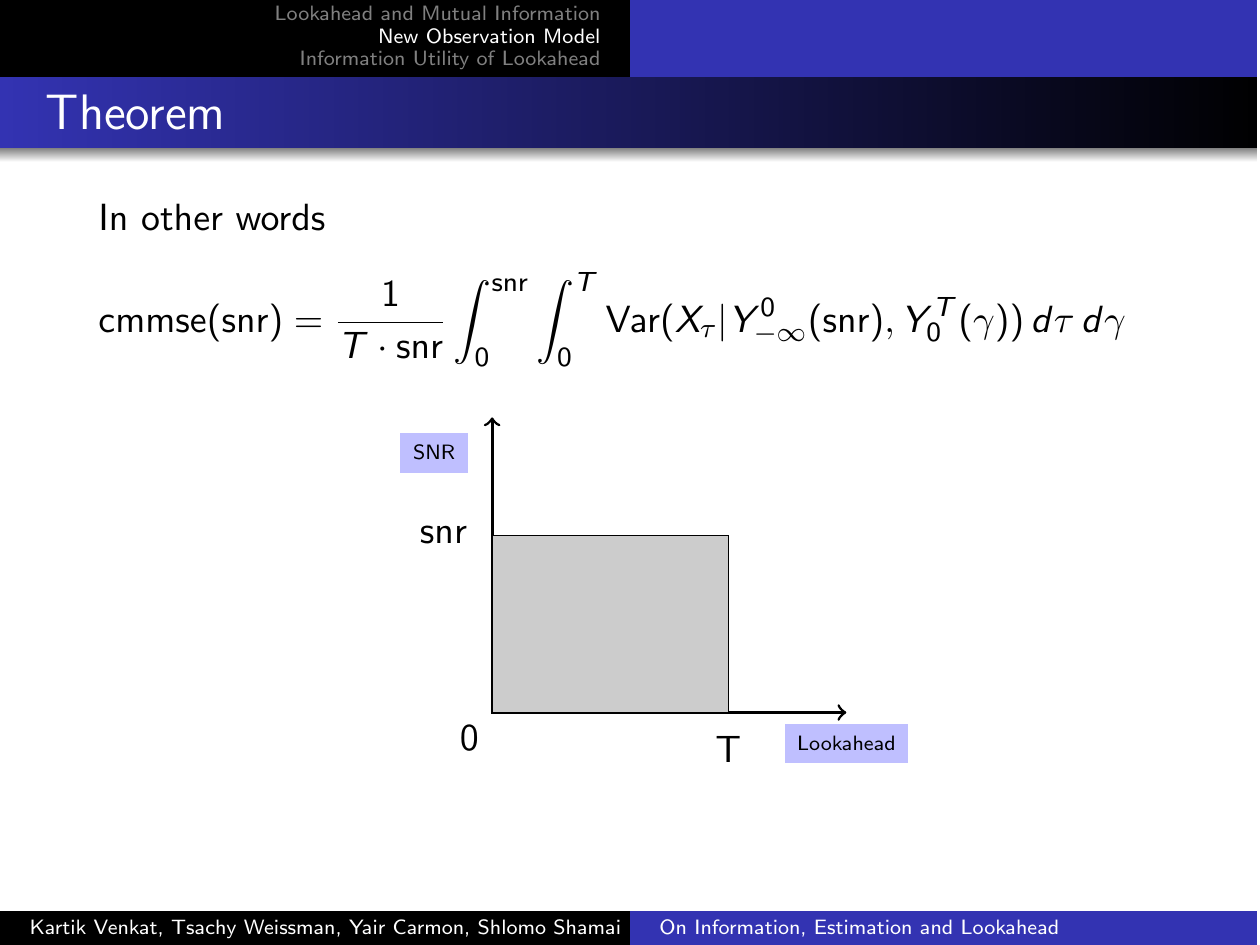}
\caption{Region in the Lookahead-SNR plane over which the finite lookahead MMSE given by (\ref{eq: function f}) is averaged over in Theorem \ref{thm: relation between f and cmmse} to yield the filtering error at level $\tsnr$.}
\label{fig: lookahead snr plane}
\end{center}
\end{figure}

It is interesting to note that the double integral over lookahead and SNR are
conserved in such a manner by the filtering error, for any arbitrary
underlying stationary process. Note that one way of interpreting this result, is to take the average of the finite lookahead mmse (under the given observation model), over a rectangular region in the Lookahead vs. Signal-to-Noise Ratio plane, as depicted in Fig. \ref{fig: lookahead snr plane}. Theorem \ref{thm:
relation between f and cmmse} tells us that for all underlying processes, this quantity is always the filtering error at level $\tsnr$. Thus, we find the classical estimation theoretic
quantity described by the causal error to emerge as a bridge between the
effects of varying lookahead and the signal to noise ratio.

Given Theorem \ref{thm: relation between f and cmmse}, it is natural to investigate whether the relation (\ref{eq: expectation identity}) can be inverted to determine the function $f$, from the causal mmse. If that were the case, then, in particular, the filtering error would completely determine the estimation loss with
finite lookahead. To pose an equivalent question, let us recall
Duncan's result in \cite{duncan} establishing the equivalence of
the mutual information rate and the filtering error.  On the other hand, by \cite{gsv2005,zakai05} the mutual information rate function $I( \cdot )$, determines the non-causal mmse.
Might the same mutual information rate function $I(\tsnr)$ can completely
determine the estimation loss for a fixed finite lookahead $d$ as well ? This question is addressed in the following section.

\section{Can I($\cdot$) recover finite-lookahead mmse ?} \label{sec: mutual
information and lmmse}
\begin{figure}
\begin{center}
\includegraphics[height=4in,width=6.5in]{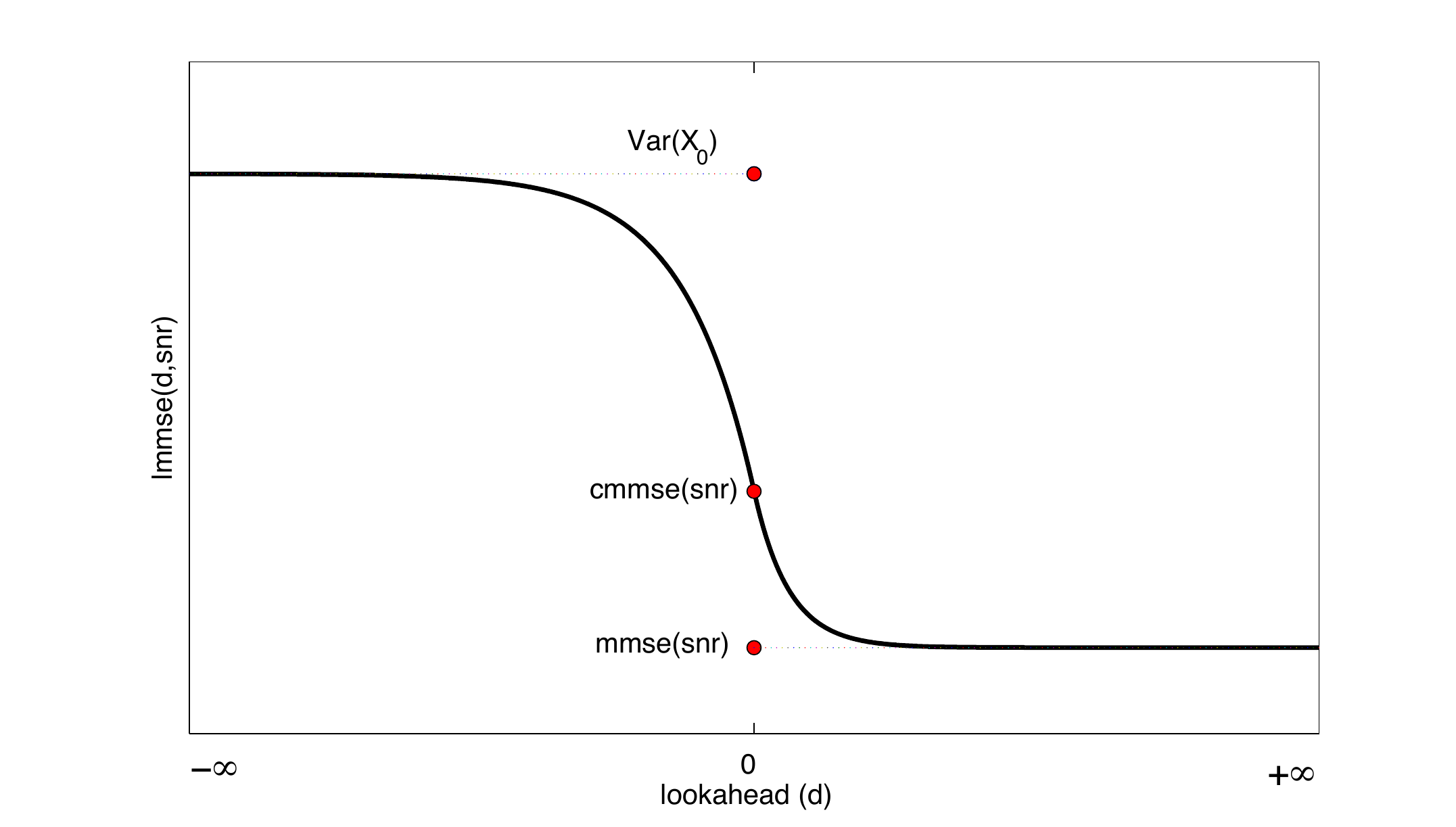}
\caption{Characteristic behavior of the minimum mean squared error with lookahead for any process.}
\label{fig: lmmse of general process}
\end{center}
\end{figure}

In Fig. \ref{fig: lmmse of general process}, we present a characteristic plot of the mmse with lookahead for an arbitrary continuous time stationary process, corrupted by the Gaussian channel (\ref{eq: channel model}) at SNR level $\tsnr$. In particular, we note three important features of the process, namely (i) the asymptote at $d = \infty$ is $\tmmse(\tsnr)$, (ii) $\lmmse(0,\tsnr) = \tcmmse(\tsnr)$ and (iii) the asymptote at $d = -\infty$ is the variance of the stationary process, $\tvar(X_0)$. Further, the curve is non-decreasing for all $d$.

From \cite{duncan} and \cite{gsv2005}, we know that the mutual information, the causal, and the non-causal mmse determine each other according to
\bea
\frac{2I(\tsnr)}{\tsnr} = \tcmmse(\tsnr) = \frac{1}{\tsnr}\int_0^{\tsnr}
\tmmse(\gamma) \,d\gamma. \label{eq: triangular relationship}
\eea
In other words, the mutual information rate function is sufficient to characterize the
behavior of the causal and smoothing errors as functions of $\tsnr$. In particular, (\ref{eq: triangular relationship}) also determines the three important features of the $\lmmse(\cdot,\tsnr)$ curve discussed above, i.e. the value \footnote{Note that $\tvar(X_0) = \tcmmse(0)$, which $I(\cdot)$ determines.} at $d=0$, and the asymptotes at $d= \pm \infty$. It is natural to ask whether it can actually characterize the entire curve $\lmmse(\cdot,\tsnr)$. The following theorem implies that the answer is negative.

\begin{theorem} \label{thm: example theorem}
For any finite $d < 0$ there exist stationary continuous-time processes which have the same mutual
information rate $I(\tsnr)$ for all $\tsnr$, but have different minimum mean squared errors with (negative) lookahead $d$.
\end{theorem}
One of the corollaries of Theorem \ref{thm: example theorem}, which we demonstrate by means of an example, is that the mutual information rate as a function of $\tsnr$ does not in general determine $\lmmse(d,\tsnr)$ for any finite non-zero $d$.

Thus, if the triangular relationship
between mutual information, the filtering, and the smoothing errors is to be extended to accommodate finite lookahead, one will have to resort to distributional properties that go beyond mutual information. To see why this must be true in general, we note that the functional $\lmmse(\cdot,\tsnr)$ may depend on the time-dynamical features of the process, to which the mutual information rate function is invariant. For example, we note the following.
\begin{observation}
Let $X_t$ be a stationary continuous time process. Define $X^{(a)}_t = X_{at}$ for a fixed constant $a > 0$. Let $Y_t$ and $Y^{(a)}_t$, denote respectively, the outputs of the Gaussian channel (\ref{eq: channel model}) with $X_t$ and $X^{(a)}_t$ as inputs. Then, for all $d$ and $\tsnr >0$,
\bea
\lmmse_{X^{(a)}}(d,\tsnr) = \lmmse_{X}(ad,\tsnr/a), \label{eq: scaling of lmmse}
\eea
where the subscript makes the process under consideration explicit. Note that for the special cases when $d \in \{0, \pm \infty\}$, the error of the scaled process results in a scaling of just the SNR parameter, i.e.
\bea
\tmmse_{X^{(a)}}(\tsnr) = \tmmse_{X}(\tsnr/a),
\eea
and
\bea
\tcmmse_{X^{(a)}}(\tsnr) = \tmmse_{X}(\tsnr/a).
\eea
For all other values of $d$, we see that the error depends on the error at a scaled lookahead, in addition to a scaled SNR level. This indicates, the general dependence of the mmse with finite, non-zero lookahead on the time-dynamics of the underlying process.
\end{observation}

One of the consequences of Duncan's result in \cite{duncan} is that the causal and
anti-causal\footnote{the anti-causal error denoted as $\text{acmmse}(\tsnr)$ denotes the filtering error for the time-reversed input process.} errors as functions of $\tsnr$ are the same (due to the mutual
information acting as a bridge, which is invariant to the direction of time).
Let $X_t$ be a continuous-time stationary stochastic process and $Y_t$ be the
output process of the Gaussian channel (\ref{eq: channel model}) with $X_t$
as input. We can write,
\bea \label{eq: relation between I cmmse and acmmse}
2I(\gamma) =  \tcmmse(\gamma) = \text{acmmse}(\gamma),
\eea
or, writing the rightside equality explicitly,
\bea
\text{Var}(X_0|Y_{-\infty}^0) &=& \text{Var}(X_0|Y_0^{\infty}).
\label{eq: forward and reverse errors}
\eea
It is now natural to wonder whether (\ref{eq: forward and reverse errors}) carries over to the presence of lookahead, which would have to be the case if the associated conditional variances are to be determined by the mutual information function, which is invariant to the direction of the flow of time. In the following we present an explicit construction of a process for which
\bea
\text{Var}(X_0|Y_{-\infty}^d) \neq \text{Var}(X_0|Y_{-d}^{\infty}) \label{eq:
forward and reverse errors are not equal}
\eea
for some values of $d$. Note that the left and right sides of (\ref{eq: forward and reverse errors are not equal}) are the mmse's with lookahead $d$ associated with the original process, and its time reversed version, respectively. Thus, mutual information alone does not characterize these objects.

\subsection{Construction of a continuous-time process} \label{sec: generate continuous time process}
\begin{figure}
\begin{center}
\includegraphics[height=2.0in,width=3.2in]{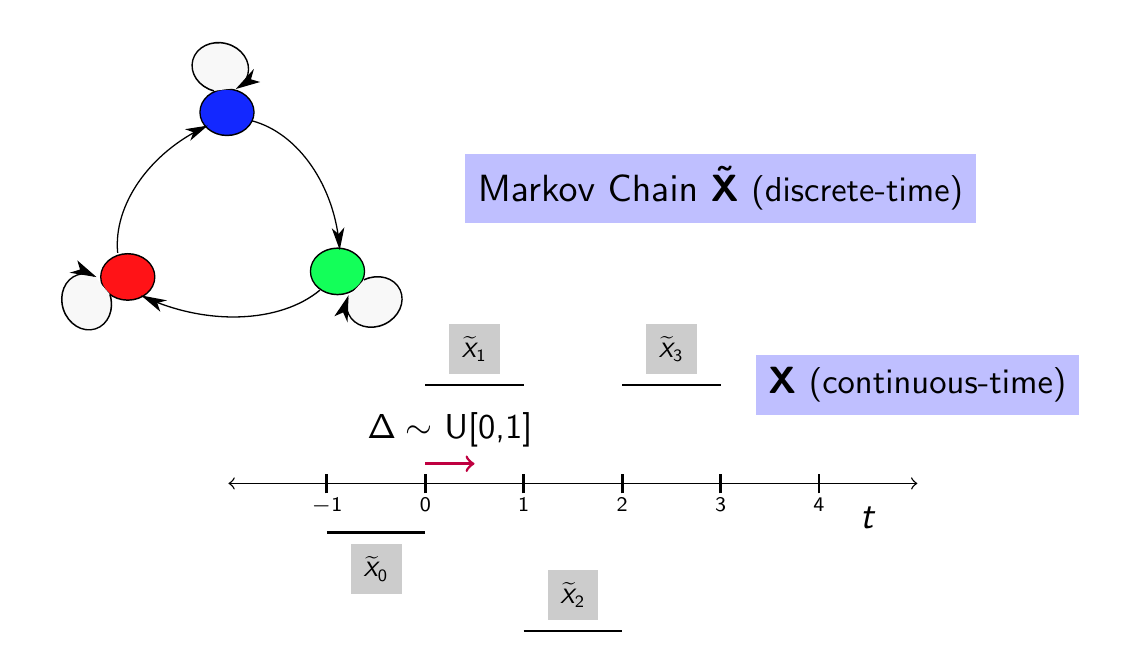}
\caption{Construction of the stationary continuous-time process $\mathbf{X}$ from the underlying discrete-time Markov Chain ${\mathbf{\tilde{X}}} $.}
\label{fig: process construction}
\end{center}
\end{figure}

In this section, we construct a stationary continuous time process from a
stationary discrete time process. This process will be the input to the
continuous time Gaussian channel in (\ref{eq: channel model}). \\
Let ${\mathbf{\tilde{X}}} = \{\tilde{X}_i\}_{i = -\infty}^{+\infty}$ be a discrete time stationary process
following a certain law $\mathbf{P}_X$. Let us define a piecewise constant
continuous-time process ${X}_t$ such that
\bea
{X}_t \equiv \tilde{X}_i \hspace{1em} t \in (i-1,i]
\eea
We now apply a random shift $\Delta \sim U[0,1]$ to the $\{X_t\}$ process
to transform the non-stationary continuous time process into a stationary one.
Let us denote this stationary process by $\mathbf{X}$. The
process $\mathbf{X}$ is observed through the Gaussian channel in (\ref{eq:
channel model}) at $\tsnr = 1$, with $\mathbf{Y}$ denoting the channel output process. This procedure is illustrated in Fig. \ref{fig: process construction}

\subsubsection{Time Reversed Process}
Consider the discrete-time process $\mathbf{\tilde{X}}^{(R)}$,
denoting the time reversed version of the stationary process
$\mathbf{\tilde{X}}$. The reversed process will, in general, not have the
same law as the forward process (though it will of course inherit its stationarity from that of ${\mathbf{\tilde{X}}}$). Let
us construct an equivalent continuous time process corresponding to
$\mathbf{\tilde{X}}^{(R)}$, using the procedure described above for the process $\mathbf{\tilde{X}}$, and label the resulting
stationary continuous-time process by $\mathbf{X}^{(R)}$. In the following
example, we compare the minimum mean square errors with finite lookahead for the processes
$\mathbf{X}$ and $\mathbf{X}^{(R)}$, for a certain underlying discrete-time Markov
process.
\subsection{Proof of Theorem \ref{thm: example theorem}, Examples and Discussions}
Define  a stationary discrete time Markov chain $\mathbf{\tilde{X}}$, on a finite
alphabet $\Xscr$, characterized by the transition probability matrix $\Pscr$.

The intuition behind the choice of alphabet and transition
probabilities, is that we would like the Discrete Time Markov Chain (DTMC) to be
highly predictable (under mean square error criterion) in the forward direction compared to the time reversed
version of the chain. Note that the time reversed Markov chain will have the
same stationary distribution, but different transition probabilities. We transform this
discrete time Markov chain to a stationary continuous time process by the
transformation described above. Because of the difference in the state predictability of the forward
and reverse time processes in the DTMC, the continuous time process thus created
 will have predictability that behaves differently for the forward and the time reversed processes and, in turn, the MSE with finite lookahead will also be depend on whether the process or its time-reversed version is considered.
\subsubsection{Infinite SNR scenario}
We now concentrate on the case when the signal-to-noise ratio of the channel is infinite. The input to the Gaussian channel is the continuous time process $\mathbf{X}$, which is constructed based on $\mathbf{\tilde{X}}$ as described above. Let us consider negative lookahead $d = -1 $. For infinite SNR, the filter will know what the underlying process is exactly, so that
\bea
\lmmse(-1,\infty) &=& \tvar(X_0 | Y_{-\infty}^{-1}) \\
&=& \tvar(X_0 | X_{-\infty}^{-1})  \\
&=& \tvar(\tilde{X}_0 | \tilde{X}_{-1}), \label{eq: infinite snr}
\eea
where (\ref{eq: infinite snr}) follows from the Markovity of $\mathbf{\tilde{X}}$. Note that the quantity in (\ref{eq: infinite snr}) is the prediction variance of the DTMC $\mathbf{\tilde{X}}$. Let $\nu: \Xscr \to \Re$ be any probability measure on the finite alphabet $\Xscr$ and $V[\nu]$ be the variance of this distribution. Let $\mu, P(\tilde{X}_{i+1} | \tilde{X}_i)$ denote respectively, the stationary distribution and the probability transition matrix of the Markov chain $\mathbf{\tilde{X}}$. Then the prediction variance is given by
\bea
\tvar(\tilde{X}_0 | \tilde{X}_{-1}) = \sum_{x \in \Xscr} \mu(x) V[P(\cdot | x)]. \label{eq: calculation of prediction variance}
\eea
In the infinite SNR setting, it is straightforward to see that
\bea
\lmmse(d,\infty) = 0, \hspace{2em} \forall d \geq 0. \label{eq: positive d infinite snr}
\eea

\begin{figure}
\begin{center}
\includegraphics[height=4in,width=6.5in]{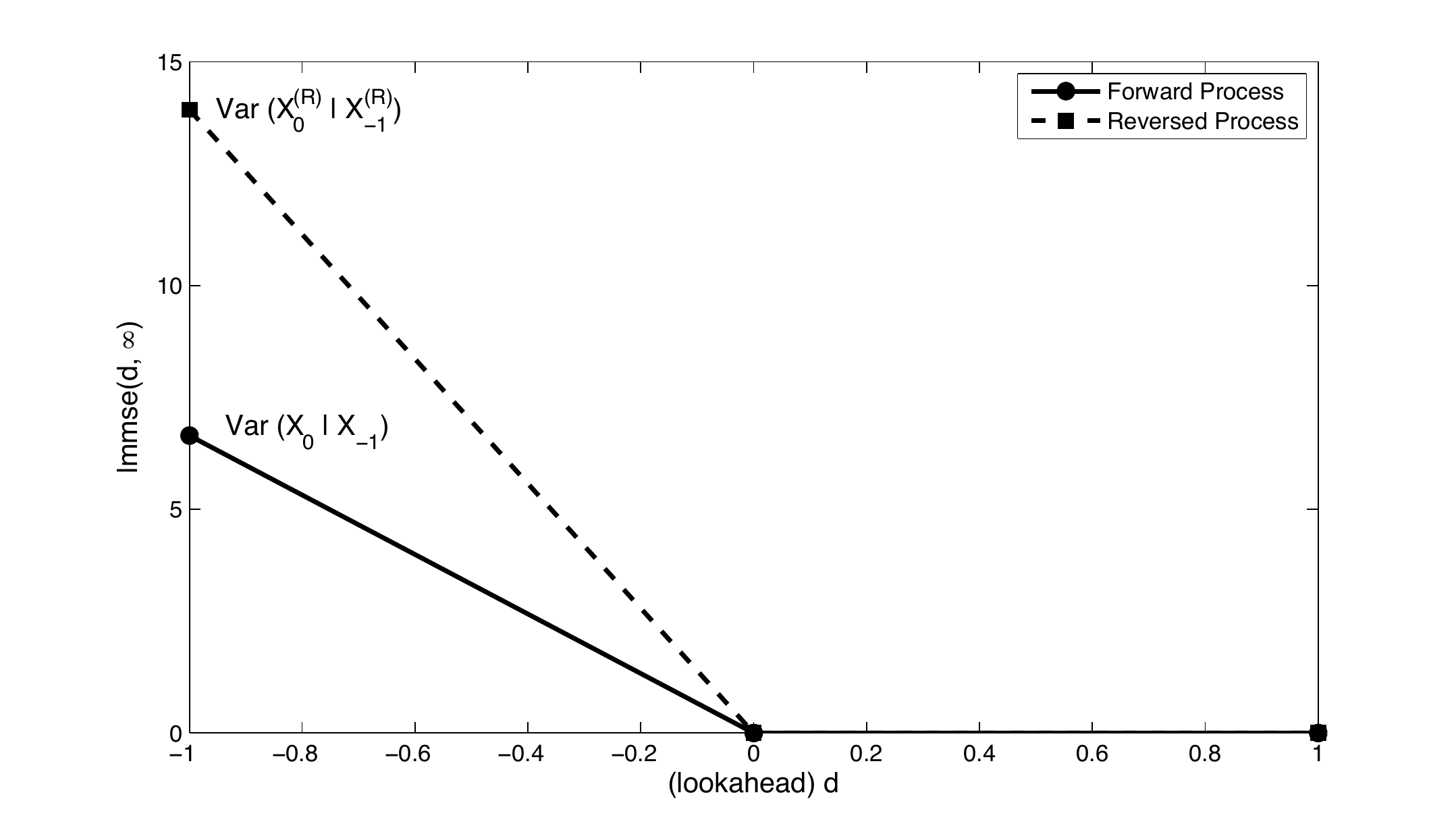}
\caption{Minimum mean squared error with negative lookahead for SNR = $\infty$ for the processes $\mathbf{X}$ and $\mathbf{X^{(R)}}$.}
\label{fig: infinite snr curve}
\end{center}
\end{figure}

Since the process $\mathbf{X}$ is constructed by a uniformly random shift according to a $U[0,1]$ law, for each $-1 \leq d \leq 0$, we have
\bea
\lmmse(d,\infty) = |d| \tvar(\tilde{X}_0 | \tilde{X}_{-1}), \hspace{1em} -1 \leq d \leq 0. \label{eq: lmmse between -1 and 0}
\eea
For fixed $d \in [-1,0]$, with probability $1-|d|$, the process $Y_{\infty}^d$ will sample $\tilde{X}_0$ in which case the resulting mean squared error is $0$ at infinite SNR. Alternately, with probability $|d|$ the error will be given by the prediction variance, i.e $\tvar(\tilde{X}_0 | \tilde{X}_{-1})$, which gives rise to (\ref{eq: lmmse between -1 and 0}).
A similar analysis can be performed for the time-reversed DTMC $\mathbf{\tilde{X}}^{(R)}$. Having found analytic expressions for the mmse with lookahead for the infinite SNR case, we show a characteristic plot of the same in Fig. \ref{fig: infinite snr curve}. Note the difference in the curves for negative lookahead arises simply due to a difference in prediction variance for the forward and time-reversed DTMC's $\mathbf{\tilde{X}}$ and $\mathbf{\tilde{X}}^{(R)}$. Since the mmse with lookahead are different for $d<0$ at infinite SNR, they would also be different at a large enough, finite signal-to-noise ratio. This completes the proof of Theorem \ref{thm: example theorem} as stated for $d<0$.

To further argue the existence of processes where $\lmmse(d,\tsnr)$ are different, also for positive $d$, we provide the following explicit construction of the underlying processes. We then provide plots of the mmse with lookahead for this process, based on Markov Chain Monte Carlo simulations. The associated plots make it clear that $\lmmse(d, \cdot)$ are distinct for both processes when $d$ is finite and non-zero.
\subsubsection{Simulations}
%Fig. \ref{fig: markov chain} represents the DTMC with transition
%probabilities as depicted.
Let $\mathbf{\tilde{X}}$ be a Discrete Time Markov Chain with the following
specifications:
The alphabet is $\Xscr = \{5, 0, -5\}$. The probability transition matrix
$\Pscr$ is given by:
\[
 \Pscr = \begin{bmatrix}
       0.6 & 0.4 & 0           \\[0.3em]
       0 & 0.2 & 0.8 \\[0.3em]
       0.875 & 0 & 0.125
     \end{bmatrix}
\]
where $\Pscr_{ij} = P(\tilde{X}_{k+1} = x_j | \tilde{X}_{k} = x_i)$. Note that $x_1 = 5, x_2 =
0, x_3 = -5$, in the above example. For this markov chain, we can compute the
prediction variance according to (\ref{eq: calculation of prediction variance}) to obtain $\tvar(X_0 | X_{-1}) = 6.6423$. The stationary prior for
this DTMC is $\mu = (0.5109, 0.2555, 0.2336)$. For the reversed process
$\mathbf{\tilde{X}}^{(R)}$, the probability transition matrix is given by
\[
 \Pscr^{(R)} = \begin{bmatrix}
       0.6 & 0 & 0.4           \\[0.3em]
       0.8 & 0.2 & 0 \\[0.3em]
       0 & 0.875 & 0.125
     \end{bmatrix}
\]

and the prediction variance is $13.9234$.\\
We performed monte carlo simulations to obtain the MSE with finite lookahead
(and lookback) for the forward and time reversed continuous-time processes thus
formed. A brief description of the simulation is provided in Appendix \ref{appendix: MCMC}.
\subsubsection{Discussion}
\begin{figure}
\begin{center}
\includegraphics[height=4in,width=6.5in]{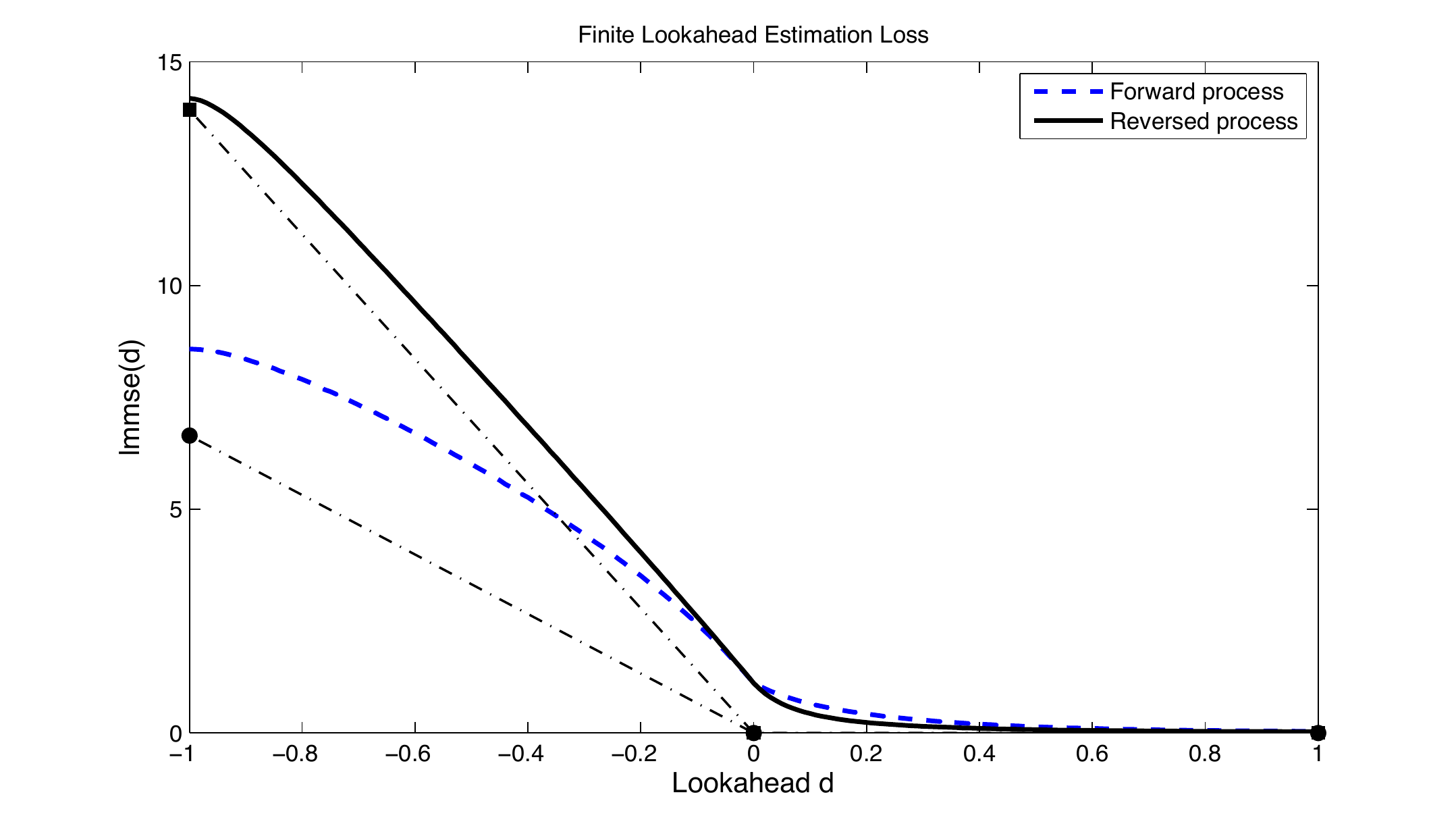}
\caption{Comparison of the Estimation Loss with finite lookahead for the
Forward and Reverse processes, $\mathbf{X}$ and $\mathbf{X^{(R)}}$ respectively at SNR=1. For reference, the infinite SNR curves (dashed) are also shown.}
\label{fig: error for forward and reverse process zoomed out}
\end{center}
\end{figure}

\begin{figure}
\begin{center}
\includegraphics[height=4in,width=6in]{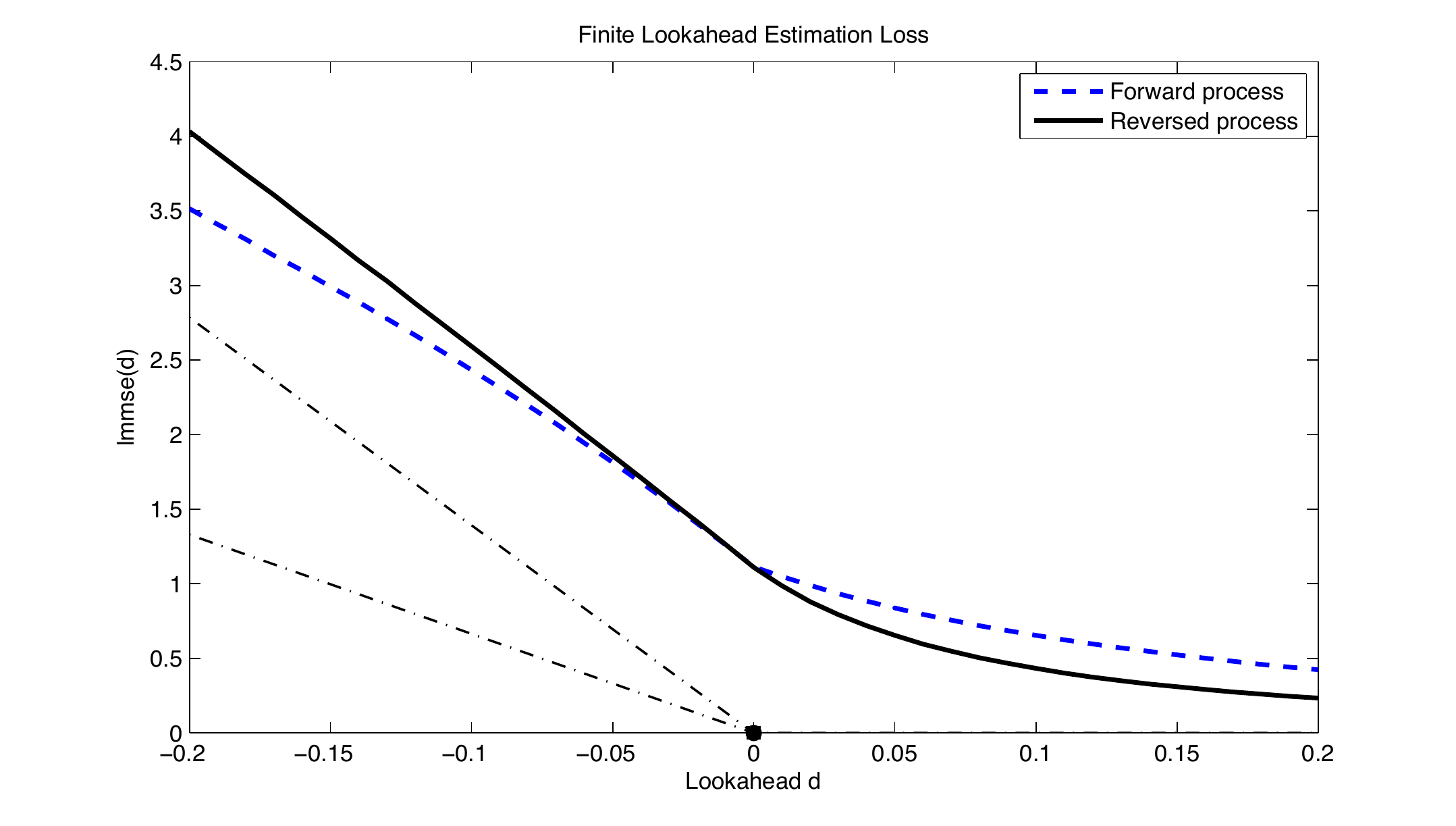}
\caption{Zoomed in version of Fig. \ref{fig: error for forward and reverse
process zoomed out}. Note that curves coincide at $d=0$, consistent with Duncan's result (\ref{eq: forward and reverse errors}).}
\label{fig: error for forward and reverse process zoomed in}
\end{center}
\end{figure}

In Fig. \ref{fig: error for forward and reverse process zoomed out}, we present
the MSE with finite lookahead and lookback for the continuous time process
$\mathbf{X}$ and $\mathbf{X}^{(R)}$ denoting the forward and time-reversed
stationary noise-free processes, respectively. From Duncan's result, we know
that the causal and the anti-causal errors must coincide. This is observed (and
highlighted in Fig. \ref{fig: error for forward and reverse process zoomed in})
by the same values for the MSE with 0 lookahead for the forward and reversed
processes. Indeed, for both positive and negative lookahead, the MSE's are
different.

Note that we know the asymptotic behavior of the MSE's with lookahead. As $d
\to -\infty$, the forward(and reverse) MSE will converge to the variance $\tvar(\tilde{X}_0)$ of the corresponding underlying
DTMC (similarly for the time-reversed chain). For $d \to \infty$, the
MSE's converge to the non-causal errors respectively (which are naturally equal for
the forward and reversed processes). Note that the forward and reversed processes have the same mutual information rate function, as well as the same spectrum. This indicates, that such measures are not capable of characterizing the MMSE with finite lookahead. 

 The above construction illustrates the complicated nature of lookahead in its role as a link between estimation and information for the Gaussian channel. While answering several important questions, it also raises new ones - Do there exist other informational measures of the input-output laws that are sufficient to characterize the estimation error with lookahead ? Such directions remain for future exploration.

\section{Conclusions} \label{sec: conclusions}

This work can be viewed as a step towards understanding the role of lookahead in information and estimation.
We investigate the benefit of finite lookahead in mean squared estimation
under additive white Gaussian noise. We study the class of Ornstein-Uhlenbeck processes and explicitly characterize the dependence of squared estimation loss on lookahead and SNR. We extend this result to the class of processes that can be expressed as mixtures of Ornstein-Uhlenbeck processes, and use this characterization to present bounds on the finite lookahead MMSE for a class of Gaussian processes. We observe that Gaussian input processes with rational spectra have the finite lookahead MMSE converging exponentially rapidly from the causal, to the non-causal MMSE. We define and obtain relationships for the information utility of lookahead. We then present an expectation identity for a generalized observation model, which presents the filtering error as a double
integral, over lookahead and SNR - of the estimation error with finite lookahead. Finally, we illustrate
through means of an example that the mutual information rate function does not uniquely
characterize the behavior of estimation error with lookahead, except in the
special cases of $0$ and infinite lookahead. Our example shows that the finite lookahead MMSE depends on features of the process that are not captured by the information rate function, or indeed, the entire spectrum. 

\appendices

\section{} \label{appendix: Lemma 1}
In this appendix we prove Lemma \ref{lemma: ed for Ornstein-Uhlenbeck process}.
Employing the continuous-time Kalman-Bucy filtering framework (cf.
\cite{LSvol1} for a detailed presentation), we obtain the following differential equation for the error
covariance $e_t$ = Var($X_t|Y_0^t$)
\bea
\frac{d e_t}{dt} &=& -2\alpha e_t - \gamma e^2_t + \beta^2, \label{eq:
differential equation for e(t) using Kalman-Bucy setting}
\eea
where $e_0 = R_X(0) = \frac{\beta^2}{2\alpha}$. We integrate over limits $0$
to $d$ and re-arrange terms to get the desired result,
\bea
\int_0^d \frac{\,de_t}{-2\alpha e_t - \gamma e^2_t + \beta^2} &=& \int_0^d\,dt \label{eq: OU kalman-bucy integral}
\\
d &=& \frac{1}{2\sqrt{\alpha^2 + \gamma{\beta^2}}}\log{\left|
\frac{-\gamma{e_d} - \alpha - \sqrt{\alpha^2 + \gamma{\beta^2}}}{-\gamma{e_d} -
\alpha + \sqrt{\alpha^2 + \gamma{\beta^2}}} \right|} - \frac{1}{2\sqrt{\alpha^2
+ \gamma{\beta^2}}}\log|\rho|,
\eea
which upon simplification yields the following expression for $e_d$
\bea
e_d &=& -\frac{\alpha}{\gamma} + \frac{\sqrt{\alpha^2 +
\gamma{\beta^2}}}{\gamma}\bigg( \frac{e^{2d\sqrt{\alpha^2 +
\gamma{\beta^2}}}\rho + 1}{e^{2d\sqrt{\alpha^2 + \gamma{\beta^2}}}\rho - 1}
\bigg), \label{eq: variance ed}
\eea
where $\rho$ is as defined as
\bea
\rho = \left|\frac{\gamma R_X(0) + {\sqrt{\alpha^2 + \gamma{\beta^2}} +
\alpha}}{\gamma R_X(0) - \sqrt{\alpha^2 + \gamma{\beta^2}} + \alpha}\right|.
\eea

% Additionally, we notice that $\hat{e}_t \doteq \tvar(X_t|Y_0^t,X_0)$ also satisfies (\ref{eq: differential equation for e(t) using Kalman-Bucy setting}), with the initial condition $\hat{e}_0$. Applying a similar integration procedure to the one described above, we find that
% \bea
% \hat{e}_d &=& -\frac{\alpha}{\gamma} + \frac{\sqrt{\alpha^2 +
% \gamma{\beta^2}}}{\gamma}\bigg( \frac{e^{2d\sqrt{\alpha^2 +
% \gamma{\beta^2}}}\hat{\rho} - 1}{e^{2d\sqrt{\alpha^2 + \gamma{\beta^2}}}\hat{\rho} + 1}
% \bigg), \label{eq: variance ehatd}
% \eea
% with
% \bea
% \hat{\rho} = \frac{\sqrt{\alpha^2 + \gamma{\beta^2}} +
% \alpha}{\sqrt{\alpha^2 + \gamma{\beta^2}} - \alpha}.
% \eea

\section{} \label{appendix: lmmse for gaussian spectrum}
In this appendix, we illustrate the computation of the MMSE with finite lookahead for a Gaussian process, whose spectrum is an equal mixture of the spectra of two Ornstein-Ulhenbeck processes, parametrized by $\alpha_i$, $i=1,2$. In this computation, we use the equations developed in Section \ref{sec: gaussian processes}. 

For simplicity, we will operate in the `s' domain for the spectra, where the region of convergence of the Laplace transform will be implicit. Recall from (\ref{eq: autocorrelation of Ornstein-Uhlenbeck process}), that the spectrum of an Ornstein-Ulhenbeck process with parameter $\alpha$ is given by 
\bea
S_{\alpha}(s) = \frac{1}{\alpha^2 - s^2}
\eea  
Thus, the spectrum of the input process $X_t$, which is a mixture of two Ornstein-Ulhenbeck processes is given by
\bea
S_{X}(s) = \frac{1}{2}\cdot\frac{1}{\alpha_1^2 - s^2} + \frac{1}{2}\cdot\frac{1}{\alpha_2^2 - s^2} \label{eq: spectrum of X}
\eea
Under the observation model in (\ref{eq: channel model}) at a fixed signal-to-noise ratio $\tsnr > 0$, the output spectrum is given by:
\bea
S_Y(s) = 1 + \frac{\tsnr}{2}\cdot \big[\frac{1}{\alpha_1^2 - s^2} + \frac{1}{\alpha_2^2 - s^2}\big] \label{eq: Sy}
\eea
Performing the Wiener-Hopf decomposition of the output spectrum in (\ref{eq: Sy}), we obtain the following spectral factors:
\bea
S^{+}_Y(s) = \frac{(s + p_1)(s + p_2)}{(\alpha_1 + s)(\alpha_2 + s)}, \label{eq: positive factor}
\eea
and
\bea
S^{-}_Y(s) = \frac{(s - p_1)(s - p_2)}{(\alpha_1 - s)(\alpha_2 - s)}, \label{eq: negative factor}
\eea
where $p_1, p_2 > 0$ are such that $p_i^2$, $i=1,2$ are solutions of $k(x) = 0$, where
\bea
k(x) = x^2 - (\alpha_1^2 + \alpha_2^2 + \tsnr)x + \frac{\tsnr}{2}(\alpha_1^2 + \alpha_2^2) + \alpha_1^2 \alpha_2^2.
\eea
Note that $S^{-}_Y(s) = S^{+}_Y(-s)$. 
Invoking (\ref{eq: def wiener-kolmogorov H(w)}), the transfer function of the optimal non-causal filter $H(\cdot)$ is characterized by 
\begin{align}
H(s) &= \frac{\sqrt{\tsnr} S_X(s)}{S^{-}_Y(s)} \\
&=  \frac{\frac{\sqrt{snr}}{2}(\alpha_1^2 + \alpha_2^2) -\sqrt{snr}s^2}{(s - p_1)(s - p_2)(\alpha_1 + s)(\alpha_2 + s)} \label{eq: formula for H}\\
&= \frac{u_1}{s - p_1} + \frac{u_2}{s - p_2} + \frac{v_1}{\alpha_1 + s} + \frac{v_2}{\alpha_2 + s}, \label{eq: partial fraction}
\end{align}
where (\ref{eq: formula for H}) follows from (\ref{eq: spectrum of X}) and (\ref{eq: negative factor}); and (\ref{eq: partial fraction}) denotes the partial fraction decomposition of the expression in (\ref{eq: formula for H}). Having derived an exact expression for the Wiener filter, we are now in a position to compute the desired quantity, i.e. $\lmmse(d,\tsnr)$. From (\ref{eq: lmmse in terms of estimation filter impulse response}) we have,
\begin{align}
\lmmse(d,\tsnr) = \tmmse(\tsnr) + \int_{-\infty}^{-d}h^{2}(t)\,dt. \label{eq: computing lmmse of gaussian spectrum}
\end{align}
From the classical result by Wiener \cite{Wiener42}, we know that the mmse of a Gaussian process is given by
\bea
\tmmse(\tsnr) = \int_{-\infty}^{\infty} \frac{S_X(j\omega)}{1 + \tsnr\, S_X(j\omega)} \, \frac{d\omega}{2\pi}. \label{eq: mmse by wiener}
\eea
Using (\ref{eq: partial fraction}) and (\ref{eq: computing lmmse of gaussian spectrum}), we are now in a position to explicitly compute the mmse with finite lookahead for the $X$ process corrupted by the Gaussian channel:
\begin{align} \label{eq: lmmse of mixture example}
\lmmse_X(d,\tsnr) = \twopartdef{\tmmse(\tsnr) + \frac{u_1^2}{2p_1}e^{-2p_1\,d} + \frac{u_2^2}{2p_2}e^{-2p_2\,d} + \frac{u_1 u_2}{p_1+p_2}e^{-(p_1+p_2)\,d}}{d \geq 0}{\tmmse(\tsnr) + C + \frac{v_1^2}{2\alpha_1}e^{2 \alpha_1\,d} + \frac{v_2^2}{2\alpha_2}e^{2\alpha_2\,d} + \frac{v_1 v_2}{\alpha_1 + \alpha_2}e^{(\alpha_1 + \alpha_2)\,d}}{d < 0},
\end{align} 
where 
\begin{align} 
C = \frac{u_1^2}{2p_1} + \frac{u_2^2}{2p_2} + \frac{u_1 u_2}{p_1+p_2} - \frac{v_1^2}{2\alpha_1} - \frac{v_2^2}{2\alpha_2} - \frac{v_1 v_2}{\alpha_1 + \alpha_2}.
\end{align}
Recall the rate of convergence of the mmse with lookahead, from causal to non-causal error, was defined as 
\bea
p_d \doteq \frac{\lmmse(d,\tsnr) - \tmmse(\tsnr)}{\tcmmse(\tsnr) - \tmmse(\tsnr)}.
\eea
For the Gaussian input process with spectrum as given in (\ref{eq: spectrum of X}), we can use (\ref{eq: lmmse of mixture example}) to compute the quantity $p_d$:
\bea \label{eq: pd for mixture example}
p_d = \frac{1}{C}\left[ \, \frac{u_1^2}{2p_1}e^{-2p_1\,d} + \frac{u_2^2}{2p_2}e^{-2p_2\,d} + \frac{u_1 u_2}{p_1+p_2}e^{-(p_1+p_2)\,d}\right]
\eea  
Clearly $p_d \tends 0$ for large $d$, in accordance with Lemma \ref{lemma: rational spectra}.

\section{} \label{appendix: OU via Ut}
Let $\hat{e}_t \doteq \tvar(X_t|Y_0^t,X_0)$, and notice that $\hat{e}_t$ satisfies the Kalman-Bucy differential equation (\ref{eq: differential equation for e(t) using Kalman-Bucy setting}),  with the initial condition $\hat{e}_0=0$. Applying a similar integration procedure to the one described above, we find that
\bea
\hat{e}_d &=& -\frac{\alpha}{\gamma} + \frac{\sqrt{\alpha^2 +
\gamma{\beta^2}}}{\gamma}\bigg( \frac{e^{2d\sqrt{\alpha^2 +
\gamma{\beta^2}}}\hat{\rho} - 1}{e^{2d\sqrt{\alpha^2 + \gamma{\beta^2}}}\hat{\rho} + 1}
\bigg), \label{eq: variance ehatd}
\eea
with
\bea
\hat{\rho} = \frac{\sqrt{\alpha^2 + \gamma{\beta^2}} +
\alpha}{\sqrt{\alpha^2 + \gamma{\beta^2}} - \alpha}.
\eea

We define the quantity,
\bea
\tau\doteq\sqrt{\alpha^2 + \gamma{\beta^2}}.
\eea
Equations (\ref{eq: cmmse for Ornstein-Uhlenbeck process}) and (\ref{eq: mmse for Ornstein-Uhlenbeck process}) can be rewritten as
\bea
\tcmmse(\gamma) = \frac{\tau-\alpha}{\gamma}
\quad,\quad
\tmmse(\gamma) = \frac{\tau^2-\alpha^2}{2\tau\gamma} = \tcmmse(\gamma)\frac{\tau+\alpha}{2\tau}.
\eea
Using the above relations, we have
\bea
\tvar(X_d|Y_0^d,X_0)&=&-\frac{\alpha}{\gamma}
+\frac{\tau}{\gamma}\left(
\frac{e^{2d\tau}(\tau+\alpha)-(\tau-\alpha)}
     {e^{2d\tau}(\tau+\alpha)+(\tau-\alpha)}\right)\\
&=& \tcmmse(\gamma) - \frac{2\tau}{\gamma}\left(
\frac{\tau-\alpha}
     {e^{2d\tau}(\tau+\alpha)+(\tau-\alpha)}\right).
\eea
Integrating, we obtain
\bea
\int_0^t\tvar(X_s|Y_0^s,X_0)ds =
t\tcmmse(\gamma)-\frac{2\tau}{\gamma}
\left[
t - \frac{1}{2\tau}\log{
    \left(\frac{e^{2{\tau}t}(\tau+\alpha)+(\tau-\alpha)}
               {2\tau}
    \right)}
\right].\label{eq: OU mmse with init conds}
\eea

Since the Ornstein-Uhlenbeck process is Markov, it satisfies $\tvar(X_t|Y_{-\infty}^t,X_0)=\tvar(X_t|Y_0^t,X_0)$ for any $t\geq0$. We may therefore use the relation in (\ref{eq: Ut MMSE stationary}) in conjunction with (\ref{eq: OU mmse with init conds}) to write an expression for the information utility of lookahead for this process:
\bea
2U(t) &=& 2{\tau}t - \log{
    \left(\frac{e^{2{\tau}t}(\tau+\alpha)+(\tau-\alpha)}
               {2\tau}
    \right)}\\
    &=& -\log{
    \left(e^{-2{\tau}t}+(1-e^{-2{\tau}t})\frac{\tau+\alpha}
               {2\tau}
    \right)}.
\eea

Since the Ornstein-Uhlenbeck process is also stationary and Gaussian, we may plug our expression for $U(\cdot)$ into (\ref{eq: lmmse from Ut MMSE}) and obtain the finite lookahead MMSE,
\bea
\lmmse(d,\gamma)&=&
\tcmmse(\gamma){\cdot}e^{-2{U(d)}}
=\tcmmse(\gamma)\left(
   e^{-2{\tau}d}+(1-e^{-2{\tau}d})\frac{\tau+\alpha}{2\tau}
   \right)\\
   &=& e^{-2{\tau}d}\tcmmse(\gamma)+(1-e^{-2{\tau}d})\tmmse(\gamma),
\eea
which recovers Lemma \ref{lemma: estimation error with finite lookahead for Ornstein-Uhlenbeck} for positive lookahead. 

\section{} \label{appendix: MCMC}
Let $\mathbf{\tilde{X}}$ be a stationary discrete time Markov chain, with known probability transition matrix $\Pscr$. Let $\tilde{Y}_{i}(\gamma)$ denote the noisy observation of $\tilde{X}_i$ corrupted via independent Gaussian noise with the signal to noise ratio of the measurement being $\gamma$. I.e.,
\bea
\tilde{Y}_i(\gamma) = \gamma \tilde{X}_i + \tilde{W}^{(i)}_\gamma,
\eea
where $\{\tilde{W}^{(i)}_\cdot\}_i$ are independent standard Brownian motions, which are further independent of the $\tilde{X}_i$'s. Let $\mathbf{X}$ denote the corresponding stationary continuous time process generated by the process described in Section \ref{sec: generate continuous time process}, and let $\mathbf{Y}$ denote the noisy process generated via the continuous time Gaussian channel (\ref{eq: channel model}) with input $\mathbf{X}$, i.e.
\bea
dY_t = X_t \,dt + \,dW_t,
\eea
where $W_\cdot$ is a standard Brownian motion independent of the $\mathbf{X}$ process. \\Define
\bea
h(\gamma_1,\gamma_2) = \tvar(\tilde{X}_0 | \tilde{Y}_{-\infty}^{-1}(1), \tilde{Y}_0(\gamma_1), \tilde{Y}_1(\gamma_2))
\eea
Let, as before
\bea
\lmmse(d) = \tvar(X_0 | Y_{-\infty}^{d}).
\eea
For $0 < d < 1$, note that
\bea
\lmmse(d) = \int_0^d h(1,d-u) \,du + \int_d^1 h(1+d-u,0) \,du. \label{eq: lmmse in terms of h}
\eea
\subsection{MCMC approach to estimating $h(\cdot,\cdot)$}
Note that $\{\tilde{X}_i, \tilde{Y}_i\}_{i \geq 1}$  for a Hidden Markov process. By using state estimation for HMP's, the following computation can be performed for any $i$,
\bea
\hat{\tilde{X}}_i = \E[\tilde{X}_i | \tilde{Y}_{1}^{i-1}(1), \tilde{Y}_{i}(\gamma_1), \tilde{Y}_{i+1}(\gamma_2)]
\eea
Also, one can observe that
\bea
\frac{1}{M} \sum_{i=1}^{M} (\tilde{X}_i - \hat{\tilde{X}}_i)^2 \to h(\gamma_1,\gamma_2), \label{eq: MCMC convergence}
\eea
as $M \to \infty$. In the simulations, we chose $M=10000$. For this value of $M$, the quantity in the l.h.s. of (\ref{eq: MCMC convergence}) was used to approximate $h(\cdot,\cdot)$, which was then used in the expression for $\lmmse(d)$ in (\ref{eq: lmmse in terms of h}) to obtain the desired values of MSE with finite lookahead, via a monte carlo approach.

Based on a similar approach, it is also possible to compute via MCMC, $\lmmse(d)$ for $-1 < d < 0$.

%\section*{Acknowledgment}



\bibliographystyle{IEEEtran}
\section*{Acknowledgment}
This work has been partially supported under a Stanford Graduate Fellowship, the National Science Foundation (NSF) under grant agreement CCF-0939370, and the Israel Science Foundation (ISF).

\end{document}